%% file: ms.tex
\documentclass[11pt]{article}

\usepackage[utf8]{inputenc}

\usepackage[backend=biber,
    style=authoryear-icomp,
    sortcites=false, 
    sorting=nyt, 
    maxbibnames=9]{biblatex}
\usepackage{geometry}
\usepackage{pdflscape}
\usepackage{afterpage} 

\usepackage{hyperref}
\hypersetup{
    colorlinks=true,
    linkcolor=blue,
    filecolor=blue,
    urlcolor=blue,
    citecolor=blue}

\usepackage{graphicx}
\graphicspath{ {./} }
\usepackage{subcaption}

\usepackage{amsmath, amssymb}
\usepackage{bm} 

\usepackage{enumitem}

\usepackage{authblk}

\usepackage{tikz}
\usetikzlibrary{bayesnet}
\usetikzlibrary{arrows}

\title{Evolution of topics in central bank speech communication}
\author[ ]{Magnus Hansson\thanks{I would like to thank Erik Hjalmarsson, Andreas Dzemski and
Marcin Zamojski for valuable comments and suggestions, as well as participants at the Centre for Finance
Research Seminar and the PhD Student Workshop at the University of Gothenburg.}}
\affil[ ]{Department of Economics at University of Gothenburg}
\affil[ ]{magnus.hansson@gu.se}
\date{\today}

\usepackage{multicol}
\usepackage{caption}
\usepackage{nonfloat}

\usepackage{rotating}

\usepackage{longtable}

\usepackage{colortbl, xcolor}
\definecolor{Gray}{gray}{0.85}
\newcolumntype{a}{>{\columncolor{Gray}}c}
\newcolumntype{b}{>{\columncolor{white}}c}

\usepackage{csquotes}

\begin{document}

\maketitle

\begin{abstract}
\input{./abstract}
\end{abstract}

\vspace{\fill} 
\noindent
\textit{JEL Classification}: C38, C55, E52, E58
\\
\textit{Keywords}: Central bank communication, Monetary policy, Textual analysis, Dynamic topic models,
Narratives
\vfill

\newpage


\section{Introduction\label{section:intro}}
\input{./intro}

\section{Data\label{section:data}}
\input{./data}

\section{Methodology\label{section:method}}
\input{./method}

\section{Results\label{section:results}}
\input{./result}

\section{FED case study\label{section:fed}}
\input{./fed}

\section{Conclusion\label{section:conclusion}}
\input{./conclusion}

\newpage

\printbibliography

\newpage

\appendix
\input{./appendix}


\end{document}

%% file: abstract.tex
\noindent
This paper studies the content of central bank speech communication from 1997 through 2020 and asks the
following questions: (i) What global topics do central banks talk about? (ii) How do these topics
evolve over time? I turn to natural language processing, and more specifically Dynamic Topic
Models, to answer these questions. The analysis
consists of an aggregate study of nine major central banks and a case study of the Federal
Reserve, which allows for region specific control variables.
I show that: (i) Central banks address a broad
range of topics.
(ii) The topics are well captured by Dynamic Topic Models. (iii) The global topics exhibit
strong and significant autoregressive properties not easily explained by financial control
variables.

%% file: intro.tex
\begin{displayquote}[Stefan Ingves, governor of the Swedish Riksbank]
I’m a weatherman, I’m a showman, and I’m an economist. I’m expected to be, and I am, a
storyteller. I tell stories about the future.\footnote{https://www.bloomberg.com/news/articles/2019-09-16/wall-street-used-to-crunch-numbers-they-ve-moved-on-to-stories}
\end{displayquote}

\noindent
Central bank communication affects financial markets
\parencite{cook1989,cochrane2002,nakamura2018} and makes monetary
policy more predictable.
Traditionally, central bank communication has had two
primary functions: To reveal private information to the public, and to affect and coordinate
financial market expectations \parencite{woodford2001,amato2002,blinder2008}.
This suggests that central
bank communication should be focused on topics closely related to monetary policy. Is this really the case?

In this paper, I give a descriptive analysis of speeches by the
major central banks connected to the Bank for International Settlements (BIS) and ask:
(i) What global topics do central banks talk about? (ii) How do these topics evolve over time?
In order to answer these questions I turn to natural language processing (NLP) to estimate
the content of speeches from 9 major central banks using Dynamic Topic Models (DTM)
\parencite{blei2006}.

The analysis shows that the central
banks talk about a broad range of topics, not all related to classic theory of central bank
communication. The estimated topics are persistent and exhibit a large
significant autoregressive effect which is robust to different model specifications. To control
for underlying variables, a case study using data from the Federal Reserve alone is
conducted.
The case study shows that topic persistence is not easily explained by controlling for regional
financial variables related to the topics. This suggests that the topics exhibit what
one might call a
\textit{narrative effect}, in which the topics are driven by economic narratives
\parencite{shiller2017}.\footnote{Although the concept of
narrative is new in economics, the idea has been present in
other fields, e.g., \textcite{sarbin1986} pioneered Narrative Psychology. In the humanities the
theory of narrative has been researched for decades, with early works such as \textcite{barthes1975}
and \textcite{bruner1991}
defining a narrative as ``an account of events occurring over time''.
For an overview see \textcite{mitchell1981}.}

The body of research relating narratives to central bank communication is growing and there are
evidence that narratives affect the economy
\parencite{nyman2018,hansen2019,ellen2019}.
These narratives could either be part of the global economy or created by the central banks themselves.
This is also consistent with the concept of
\textit{gradualism} in which central banks tend to gradually change the interest rate rather than
have large jumps. For instance, from 2001 to 2003 the Fed used gradualism to reduce the interest
rate by 550 basis points over a sequence of thirteen cuts.\footnote{This narrative was
discussed by Mr Ben Bernanke, member of the Board of Governors of the US Federal Reserve System,
at his speech \textit{Gradualism} at an economics luncheon co-sponsored by the Federal Reserve Bank
of San Francisco (Seattle Branch) and the University of Washington, Seattle, 20 May 2004.}
Gradualism is founded on the theory of uncertainty in policy making, in which
policymakers are inclined to gradually introduce a policy when its effect on the economy is
ambiguous \parencite{brainard1967}. Similarly, central bank communication can introduce
narratives that gradually prime the public for future policy changes, such as changes in
financial market regulation or an introduction of central bank digital currency (CBDC).
At present, many of the central banks are investigating a potential
introduction of digital money and many of them are actively communicating on the matter
through speeches and reports \parencite{cbdc1}.\footnote{The ECB, Bank of Japan, Sveriges
Riksbank, Swiss National Bank, Bank of England and the Fed, are actively
investigating and reporting on CBDCs.}
Thus, central banks are using verbal communication to inform and prepare the public for future
structural change. This is an undocumented way for central banks to use communication in an active
manner to make sure that monetary policy and regulatory changes are achieved as expected.
This indicates that central bank speeches might address a broader range of
topics than expected by the traditional view of central bank communication.

The two most common ways for central banks to communicate are: (i) Written disclosures of
meeting minutes and reports. (ii) Speeches.
Since the 1990s central bank communication strategies have undergone a transformation: Going from opaque
secrecy, to greater transparency, to actively using communication as a tool for monetary policy
\parencite{woodford2005,blinder2008,blinder2018}.
One of these changes is a large increase in the number of speeches held in the overall communication
strategies.
This has led to a great growth in text data related to central banks, which now can be analyzed
by recent machine learning techniques.
Text data is multidimensional and rich in information.
Using advances in computational linguistics, it is possible to reduce the
dimensionality of the data and use it in economic analysis.
The number of speeches from the central banks affiliated with the BIS, treated in this paper,
was 119 in 1997, whereas in 2019 it was 423.
Thus, the volume of the oral communication has increased greatly.
Central bank speeches, in comparison to
announcements, are richer in information, greater in number, significantly longer, and addresses a
larger variety of topics.
Therefore, speeches are ideal data to analyze the content of central bank communication.

Some research has previously been conducted analyzing central bank speeches.
\textcite{jansen2005} show that
speeches from the ECB affect the volatility of the euro-dollar exchange
rate. \textcite{andersson2006} study speeches from the Swedish Riksbank and find that the
speeches affect market prices, and that the market react stronger to communication from
the head of the Riksbank.
\textcite{born2014} show how sentiment of central bank speeches about financial stability
have a significant effect on market returns and volatility. The two closest
studies to this one, which also rely on topic modelling, are \textcite{hansen2017} and
\textcite{armelius2020}.
\textcite{hansen2017} study central bank transparency using topic modelling in an event study
around 1993, the year the Fed started to release the FOMC meeting transcripts.
\textcite{armelius2020} study spillover effects in sentiment from central bank speeches and show
that cross-country effects affect both central bank communication as well as macroeconomic
variables, where the Fed has a unique influence in creating sentiment spillover effects.

When the interest rate is close to the efficient lower bound (ELB),
central bank communication is of increased importance, and forward guidance may be the main policy
tool \parencite{blinder2008}. At these times, the public's expectations of the central bank's future
policy is crucial, and indicates that central bank communication might be weighted towards forward
guidance. Yet, the results of this paper suggest that
the content of central bank communication is broad, also at times when the interest rates are close
to the ELB. This is consistent with the research showing that the general public is
either not targeted or affected by central bank communication
\parencite{kumar2015,lamar2019,coibion2020}, and the fact that trust in central banks is relatively low
\parencite{hayo2014,eb92}.\footnote{According
to the Eurobarometer survey, public trust in the ECB is low \parencite{eb92}, but increases as
communication from the ECB increases \parencite{hayo2014}. Public trust in the ECB has fluctuated
during the last decade. At its lowest in 2014 it was $31\%$, and $42\%$ at its highest
in 2019 \parencite{eb92}.}
\textcite{blinder2018} predicts that ``central banks will keep trying to communicate with the
general public, as they should, but for the most part, they will fail''.
Drawing from the literature of narrative economics, \textcite{haldane2018} argue that
effective communication needs to be simple, relevant, and story-based and that central bank
communication fail at all three parts, making it inaccessible for the society in general.

This paper contributes to the literature in several ways. First, the paper provides a comprehensive
dynamic analysis, not previously conducted, describing the evolution of the content of central bank
communication.
Second, the paper introduces the application of Dynamic Topic Models
to the field of finance and economics.
Third, I make inference about the
autoregressive properties of the estimated topics, and show that there is strong persistence in
the content, which is not explained by controlling for underlying financial variables.
Finally, the paper draws connections between central bank communication, topic
modelling, and narrative economics, yielding ideas for further improvements in central bank
communication strategies, and suggestions on how topic modelling could be applied to
narrative economics.

The rest of the paper is organized as follows: Section \ref{section:data} describes the data;
Section \ref{section:method} outlines the methodology; Section \ref{section:results}
presents the main findings; in Section \ref{section:fed} a case study investigating the persistence of
topics in speeches from the Federal Reserve is conducted, controlling for regional financial
variables; and Section \ref{section:conclusion} concludes the paper.

%% file: data.tex
The central bank speech data  was scraped from the BIS website.\footnote{The data was scraped using the Request and
Beautiful Soup Python libraries.}
14,423 central bank speeches were collected from 113 institutions, over the time period
1997 through 2020. The speeches consist of English sentences, meaning that the speeches are
transcribed, and when needed translated, into only content and not including verbal sounds or utterances
without meaning. This leads
to a simpler preprocessing as the text does not contain dis-fluencies, such as fragments of words
or filled pauses.
\textcite{armelius2020} were first to use the BIS data source, using a shorter time period,
and to my knowledge their
paper is the only previous time the data source has been used.

The dataset was constrained to include global institutions with more than 200 speeches
over the sample period. This criteria was chosen in order to select a sufficiently homogenous
global and talkative subsample. Thus, local central bank branches,
such as the Bank of Spain, were excluded from the dataset. The final data for the analysis
consist of 7,379 speeches from 9 different central banks: Bank of Canada, Bank of England, Bank of
Japan, Central Bank of Norway, ECB, Fed (including speeches from the New York Fed), Reserve
Bank of Australia, Sweden's Riksbank, and Swiss National Bank.

The preprocessing of the text data follows standard methodology \parencite{gentzkow2019}. The data
were first transformed from pdf format to text format.\footnote{The Textract Python library was used for
this task.}
Each document was split into lower case tokens (words), removing punctuation, numbers and web
links.
Headers and footers of the documents were removed, together with reference lists. A
common list of stop words was applied to filter out words of little importance to the topic
modelling. Through lemmatization the tokens were transformed into dictionary form, e.g.,
\textit{banks} becoming \textit{bank}.
Bigrams, i.e, sequences of two adjacent tokens, and trigrams, i.e., sequences of three adjacent
tokens, were
created of commonly followed tokens, such as \textit{central bank} and \textit{real interest rate}.
Extreme tokens, appearing less than 20 times in the corpus or in
more than 50\% of the documents, were also filtered out to reduce dimensionality and make
interpretation simpler by avoiding topics to have the same top words in the generating
distributions.\footnote{The degree of filtering
has in this paper been determined by a grid search over topic coherence \parencite{newman2010},
using the full sample and an LDA model.}
Table \ref{table:summary} shows the data dimensionality reduction at each step in the
preprocessing.
After preprocessing, the data consist of 4,280,706 tokens and the vocabulary (alias dictionary) of 20,697 unique
tokens.\footnote{A kernel density estimation shows that the length of the speeches and the
vocabulary are smoothly distributed over the corpus, indicating that the speeches are
not clustered but continuously distributed.}

\begin{table}[h]
    \centering
    \input{./summary_table_mod}
    \caption{Data dimensionality reduction of each preprocessing step.}
    \label{table:summary}
\end{table}

The control variables for the Fed case study were collected from the Wharton Research Data Service
(WRDS). The data include: 1 year US treasury bond yields, US inflation, the S\&P 500 Index returns, and the
CBOE Volatility Index (VIX).
The S\&P 500 Index data and the VIX data have been down-sampled into quarterly observations by choosing
the maximum values in each quarter. The maximum values were chosen to preserve as much of the
variance in the data as possible.

%% file: summary_table_mod.tex
\scalebox{0.8}{
\begin{tabular}{a|b|a|b|a|b}\hline\hline
&Raw text&Remove stopwords&Lemmatization&Bigrams and trigrams&Filter extremes\\
\hline
Total words&22,762,644&11,903,054&11,737,316&9,908,456&4,280,706\\
Unique words&66,735&58,478&53,116&65,282&20,697\\
\hline\hline
\end{tabular}
}

%% file: method.tex
In order to analyze the central bank speech data,
I use Dynamic Topic Models (DTM) \parencite{blei2006} together with Autoregressive (AR) models.
DTM has not, to my knowledge, previously been applied in the field of finance and economics.
DTM let the topics dynamically change
over time, meaning that the word distributions that define the topics are dynamic. This
allows the researcher to study the time evolution of the latent dynamic topics
discovered by the model. It also
improves upon transparency when using the estimated topics in time
series modelling, since the researcher can verify homogeneity of the topic distributions over time.
One can thus identify whether a topic is about the same subject throughout the sample period.

Latent Dirichlet Allocation (LDA) \parencite{blei2003} has become the standard
topic model in the applied literature, building on Latent Semantic Indexing (LSI) introduced by
\textcite{deerwester1990}, and later extended by \textcite{papadimitriou1997} and
\textcite{hofmann1999}.\footnote{The LDA
algorithm has been further developed:  HDP (Hierarchical Dirichlet Process) takes
an hierarchical topic structure into account to find the number of topics \parencite{teh2004},
and CTM (Correlated Topic Models) \parencite{blei2007} assumes correlation between topics.}
These methods have become increasingly popular in finance and economics, for an overview of
NLP in economics see \textcite{gentzkow2019}.\footnote{Further,
efforts have been made to write software to make natural language processing and  topic modelling
more easily accessible, e.g., see \textcite{mccallum2002}, \textcite{bird2009} and
\textcite{rehurek2010}.}
Like the LDA model, DTM are a set of generative probabilistic models for
discrete data, popular for text, that introduce a time dimension to the LDA framework and let
topics evolve over time.\footnote{Dynamic
Topic Models (DTM) are also called Dynamic LDA (D-LDA) and Sequential LDA (SLDA).}
The DTM are unsupervised and use a
bag-of-words structure, meaning that the order of the words do not matter.
However in contrast to the static LDA model, the order of the
documents do matter.
The dynamic model addresses the static assumption of the LDA model by updating the distributions
parameters for each time slice, which in this paper is done annually.
This is done by introducing a state space model using a logistic
normal distribution.
In the model, each document is generated from a mixture of topics and each topic
is generated from a mixture of words from the vocabulary.

Given a model with $K$ topics, $D$ documents, and a vocabulary with $V$ terms,
let $\beta_{t, k}$ be a $V$-dimensional vector representing topic $k$ at time $t$,
where $t=1,\ldots , T$ and $k=1, \ldots , K$. $\beta_{t,k}$ evolves with a Gaussian random walk,
$\beta_{k,t} \vert \beta_{k,t-1} \sim \mathcal{N}(\beta_{k,t-1}, \sigma^2 I)$,
meaning that the word distribution over topics change over time. Furthermore, let $\alpha_t$ be
a $D$-dimensional mean parameter vector of the logistic normal distribution for the topic proportions,
following a Gaussian random walk, $\alpha_t \vert \alpha_{t-1} \sim
\mathcal{N}(\alpha_{t-1}, \delta^2 I)$. A set of topic models are sequentially chained
together and the generative process for time slice $t$ of a sequential corpus is as follows:

\begin{enumerate}
    \item Draw topics distributions over dictionary $\beta_{k,t} \vert \beta_{k,t-1} \sim
        \mathcal{N}(\beta_{k,t-1}, \sigma^2 I)$.
    \item Draw mean parameters of document distributions over topics $\alpha_t \vert \alpha_{t-1} \sim
        \mathcal{N}(\alpha_{t-1}, \delta^2 I)$.
    \item For each document:
        \begin{enumerate}[label=(\alph*)]
            \item Draw $\eta \sim \mathcal{N}(\alpha_t, a^2 I)$.
            \item For each word position $n \in N_d$:
                \begin{enumerate}[label=(\roman*)]
                    \item Draw topic $Z \sim Mult(\pi(\eta))$.
                    \item Draw word $W_{t,d,n} \sim Mult(\pi(\beta_{t,z}))$.
                \end{enumerate}
        \end{enumerate}
\end{enumerate}
Here $\pi(\beta_{k,t})_w = \frac{\exp (\beta_{k,t,w)}}{\sum \exp (\beta_{k,t,w)}}$
maps the multinomial natural parameters to the mean parameters.\footnote{Note that the
process is similar to that of LDA \parencite{blei2003}. However in LDA, the topics and the topic
proportions would have been sampled from the static Dirichlet distribution, which is the conjugate
prior to the Categorical distribution, i.e., a generalization of the Bernoulli distribution or a
special case of the Multinomial distribution (one draw instead of many), which simplifies the
estimation process, and allow for efficient use of Gibbs sampling. For a practical overview of LDA
see \textcite{griffiths2004}.} From a practical point of view, one does not generate the corpus, but
rather backs out the underlying latent distributions, given a corpus, with variational Bayesian
inference.\footnote{\textcite{blei2006} discuss both Variational Kalman Filtering, as well as
Variational Wavelet Regression.}
The standard deviations of the Gaussian random walks are not estimated but set to a fixed value
given by the implementation in \textcite{blei2006}, which is used in this paper together with a
Python wrapper \parencite{rehurek2010}. To not estimate these hyperparameters is standard in the literature
but nevertheless a limitation of the methodology since it restricts to what degree the topics can
change over time.

In a practical way the model can be understood to yield two kinds of results. First, the model
outputs a set of $K$ $V$-dimensional topic distributions for each time slice $t$. These topic
distributions are functions of the vocabulary and define
the $K$ estimated topics in the model. In a topic's probability
distribution, each word in the vocabulary at each time slice is assigned a probability defining how
likely that word is to be drawn from that topic at time $t$.
Therefore, the most probable words of a topic's distribution constitute the theme of that topic. The
topic names are manually labelled by these themes. By
tracking the changes in a topic's distribution across time it is possible to study how the
topic evolves.
An example can be seen in Table \ref{table:dtm_supervision} in Section
\ref{section:results_topics} for the estimated topic about supervision and regulation.
Second, with a trained model each document in the corpus can be
assigned a static $K$-dimensional distribution of topics, where
each topic is assigned a probability. This means that each document is classified with a set of
topic probabilities describing how likely the document is to have been generated from each of the topics. By
classifying the documents and averaging the distributions on a monthly or quarterly basis, it is
possible to track the evolution of topics discussed in the documents over time, as seen in Figure
\ref{fig:stackplot} in Section \ref{section:results_topics}. Thus, by the output of the model we
are able to study both the evolution of topics discussed throughout the sample period,
as well as the terminology-evolution of the topics' distributions.

The progression of topics discussed and the within topic change are both functions of the
underlying corpus. If certain significant global events are happening on the financial markets we
would expect a topic about financial markets to include the contemporary relevant language
and have a higher probability of being addressed. However, the emerging literature of
narrative economics \parencite{shiller2017} suggests that there are additional variables affecting
the development and spread of topics.
Narrative economics can be an explaining theory of what is driving the unexplained
persistence in the model.
Narratives in the global economy, or narratives created by the central banks themselves, can be
contributing factors to what words are likely to appear in the
topics' distributions in each time slice, as well as what topics are likely talked about in each
time slice.

Choosing the number of topics, $K$, is a non trivial and highly researched area.
In the DTM framework, the number of topics are assumed to be known, and therefore needs to be
specified beforehand.\footnote{In contrast to DTM the Hierarchical Dirichlet Process (HDP) \parencite{teh2004} is a
topic model built to uncover the number of underlying topics.}
A common way to determine the number of topics is to evaluate the model according to model
perplexity \parencite{wallach2009}
in which the inverse of the geometric mean of the per-word likelihood is evaluated. Another way is to
evaluate the topic coherence, introduced by \textcite{newman2010}, which addresses the issue of human
interpretability of the topics. If a topic is to make sense for humans, the top words
contributing to the topic distribution needs to be semantically close to each other.
This can be measured numerically with different coherence measures.
In this paper the number of topics was chosen by a grid search evaluating the topic coherence
measure implemented by \textcite{roder2015} using an LDA model on the whole corpus.

Common problems in unsupervised topic modelling are residual topics without a clear
meaning and topics that are too similar to be distinguished from each
other in a meaningful way.
To combat these problems some researchers choose a large amount of topics, e.g., 100, and discard the irrelevant
topics. Another method would be to gather similar topics together and manually classify them as one.
Although these are common approaches, it can lead to overfitting and ambiguity.
In this paper I was able to preprocess the data in such a way that meaningful and distinct topics
were obtained when optimizing over the coherence score without resorting to increasing the number
of clusters in a dramatic way, or any other human involvement.

%% file: result.tex
\subsection{Central bank topics\label{section:results_topics}}
The hyperparameter optimization subject to the coherence measure yields 29 topics in total.
After the DTM estimation, the topics were categorized into 12 local and 17 global topics,
depending on the themes of the most likely words in the
estimated probability distributions. The local topics are
region specific, meaning that the highest-probability words belonging to these topics
are related to the corresponding regions. For example, top tokens
associated with the topic about the Swedish Economy are \textit{Sweden}, \textit{Riksbank}, and
\textit{Swedish}, and words relevant to the Swedish economy, such as \textit{Krona}
and \textit{Repo rate}. Furthermore, the probability that a local topic is mentioned in an
international setting is low, e.g., the Swedish topic is mainly talked about by the Swedish
Riksbank, hence this is another way to verify and distinguish local from global topics.

\begin{figure}[h]
    \includegraphics[width=1\textwidth]{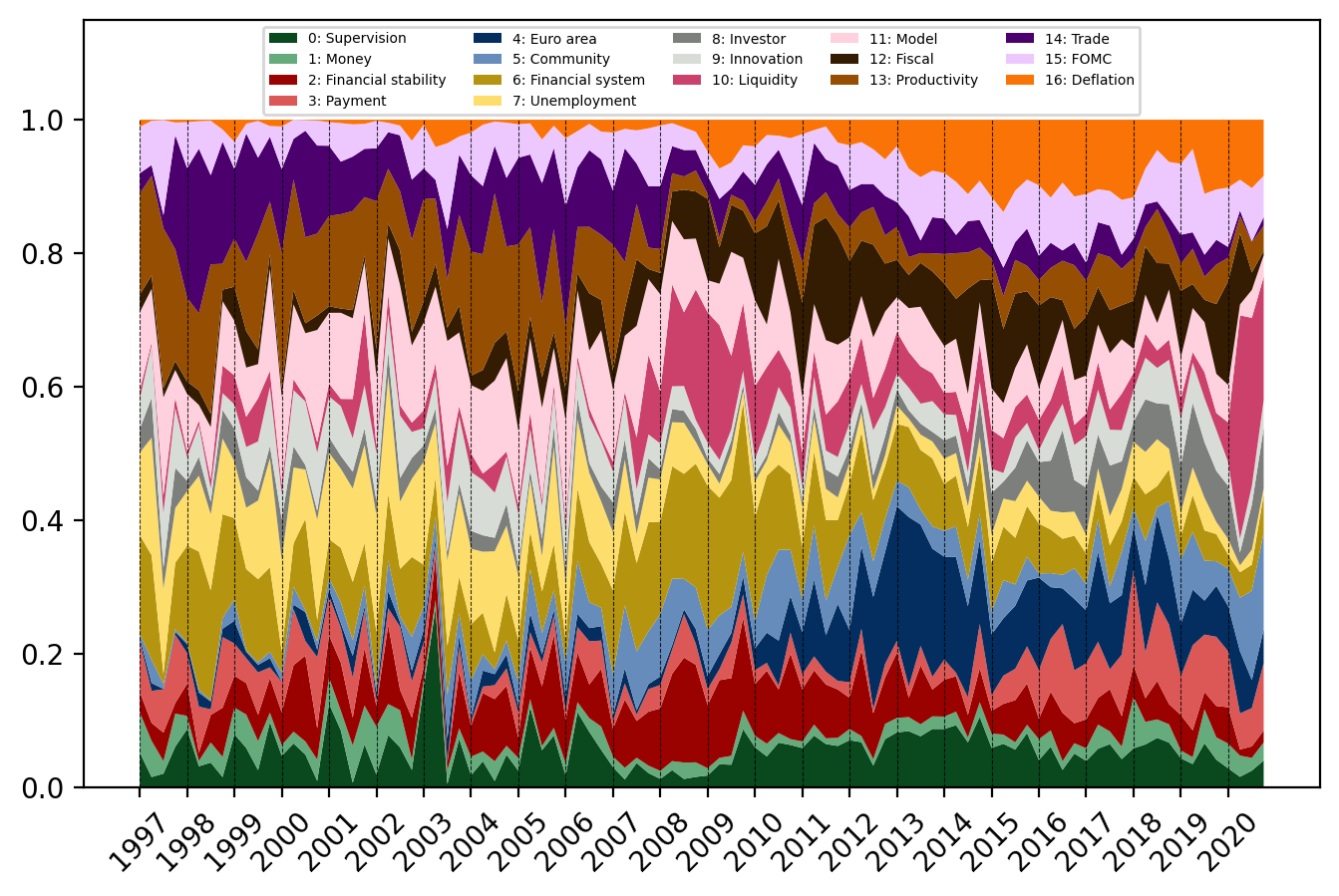}
    \centering
    \caption{Plot of the average normalized global topic probabilities of the classified documents
    for each quarter.}
    \label{fig:stackplot}
\end{figure}

Figure
\ref{fig:stackplot} shows the average normalized probability distributions for the
global topics for each quarter in the sample period, given by the classified documents in the corpus.
In each quarter all speeches are classified, their probability distributions averaged (and
normalized), and plotted.
As seen in the figure, the model captures 17 global topics that are continuously present throughout
the sample period, and thus gives an overview of  what central banks talk about. The
topics are, as expected, primarily about central bank issues, such as monetary policy. However, the
analysis shows that the topics are not mainly about coordination of financial market
expectations, but addresses a rather broad set of themes.
This suggests that central bank communication is not consistent with the traditional
view that it is fundamentally used to reveal private information to the public and to affect and
coordinate financial market expectations. Figure \ref{fig:stackplot} shows that the central
banks' communication has been seemingly diverse throughout the sample period, targeting a wide
range of global topics, including payment
systems, small business communities, innovation and technology, economic modelling, and
productivity. This holds true also in times were global interest rates have been close to the ELB.
One can argue that the
topics discussed are in the long run helping the central banks to communicate monetary policy. By
informing the public regarding matters such as payment systems or innovation and technology,
central banks can build up narratives and prime the economy for future monetary policy paradigm
changes. Therefore the communication can be seen as indirectly related to the traditional
definition of central bank communication.

Some topics have seemingly constant mass over the sample period,
e.g., the topic regarding supervision and regulation.
Other topics exhibit clear trends. After the global financial crisis in 2008,
one can see a clustering in topics associated with the financial system, financial stability,
and liquidity. In contrast, other topics exhibit steady upward trends over time.
The topic about the payment system has a positive recent trend, which is in line with
contemporary technical advancements in the area. Various countries see clear increases in the amount
of digital transactions and many central banks are investigating possibilities for CBDCs.

An advantage of using Dynamic Topic Models, compared to static topic modelling such as
LDA, is to be able to follow the topics dynamically through the time dimension.
Table \ref{table:dtm_supervision} shows a sample of the year-by-year evolution of the probability distribution
of the topic related to supervision and regulation. The table shows
words with the highest probability of being drawn from the topic from years in the beginning and
the end of the sample period. One can see that the topic is apparently homogenous
over time, indicating that in the beginning of the sample, as well as in the end, the topic is
about precisely supervision and regulation. This is important in order to be able to draw more rigid
conclusions from further
econometric analysis using the estimated topics. If one concludes that there is persistence in a
topic it is appropriate to know the content of this topic through time.
Furthermore, the dynamic analysis allows for
within-topic investigation of the vocabulary distribution.
Even though the topic is stable one can see that the vocabulary has
changed over time and the probability of other words have increased within the distribution.
In Table \ref{table:dtm_supervision}, the token \textit{stress test} has a higher
probability in the end of the sample period. However, the token \textit{capital requirement} is present as a
top word both in 1997 as well as in later years. Note that the persistence in the word-probability
distribution of topics is to some extent built into the model. The standard deviation in the
model controls the speed at which the topics can evolve on an annually basis.
The lower the standard deviation the lower is the variation in topics over time, and with a
standard deviation of zero the model becomes static.

\begin{table}[h]
    \centering
    \input{./topic_0_mod}
    \caption{Year by year evolution of the probability distribution associated with the
    topic about supervision and regulation. Sorted probabilities in parenthesis.}
    \label{table:dtm_supervision}
\end{table}

\begin{figure}[h]
    \includegraphics[width=1\textwidth]{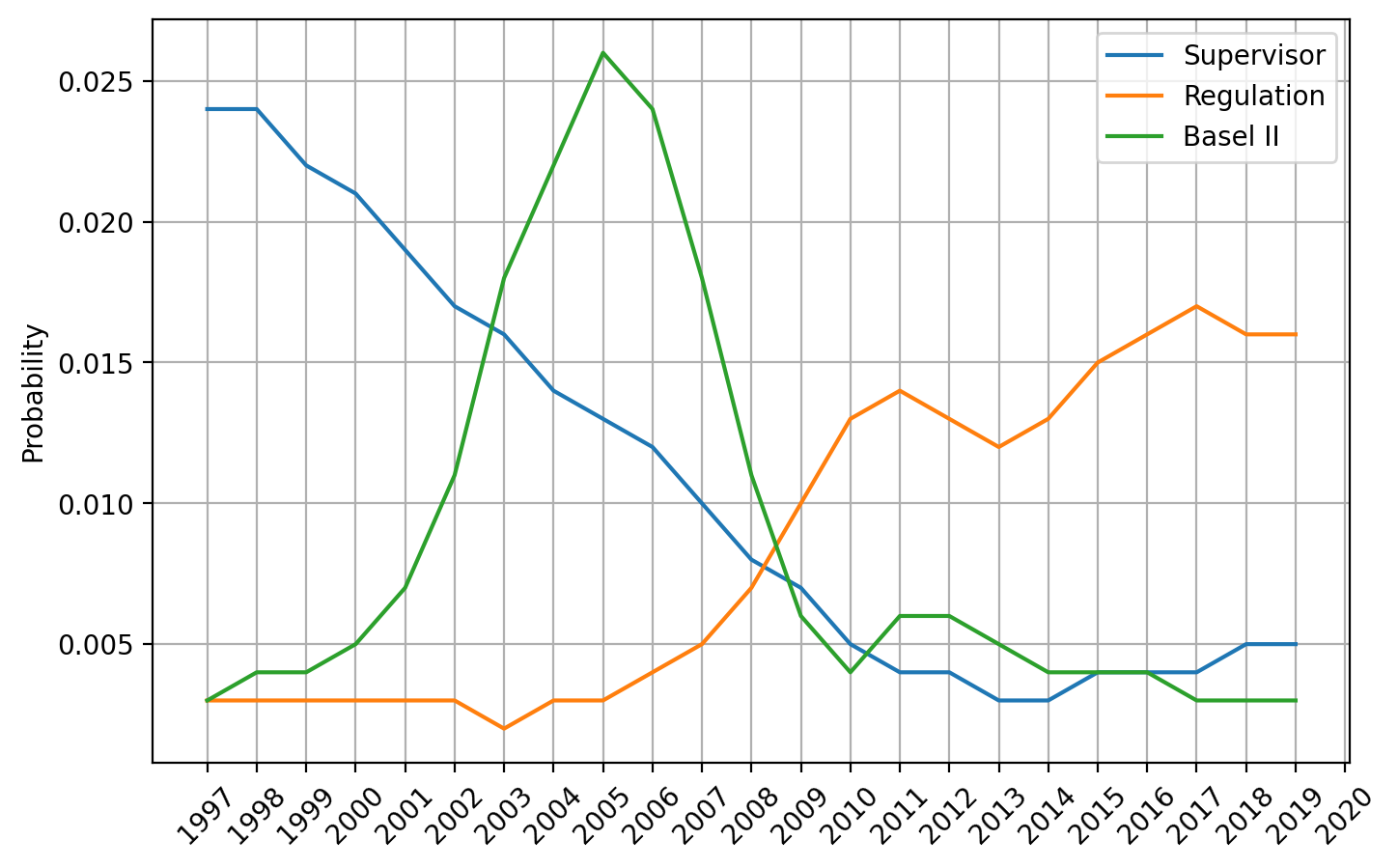}
    \centering
    \caption{Plot of the average probabilities of the tokens; \textit{Basel II}, \textit{Supervisor},
    and \textit{Regulation}, associated with the topic about supervision and regulation.}
    \label{fig:supervision}
\end{figure}

Figure \ref{fig:supervision} shows the probability, throughout the sample, of the tokens; \textit{Basel II},
\textit{Regulation}, and \textit{Supervisor}, from the topic about supervision and regulation.
One can see that there has been a shift in
language from using the word \textit{supervisor} to using the word \textit{regulation}, and the
curves intersect around 2008. This is in line with the fact that the global financial
crisis led to new regulations in the global financial industry.
Furthermore, the token \textit{Basel II} constitutes a bell shaped curve in Figure
\ref{fig:supervision}, with its highest probability just after the Basel II Accord was published in
June 2004.
The bell shaped pattern matches the theory of epidemiology of narratives discussed in
\textcite{shiller2017}. A narrative starts,
it grows, then peaks, and finally declines. Here the underlying narrative is easy to understand, since
it depends on the regulations and guidelines issued by the Basel Committee on Banking and
Supervision.

\textcite{gentzkow2019} emphasize the value of human cross-checking when the subsequent results of
natural language processing are used beyond prediction, in descriptive or statistical
analyses.
An auditing of a subsample of twenty to thirty documents will usually give an idea if the model
captures the relevant information in the corpus.
In topic modelling, it is important to validate that the topics are doing a good job in explaining the
documents they are supposedly generating. Documents are generated from a mixture of topics. However,
some documents have likely been generated from one topic alone,
and are thus excellent candidates for manual investigation.
The speech ``Implementing Basel II -- choices and challenges'' by Ms Susan Schmidt Bies at the Fed
is the speech in the corpus with the highest probability (99\%) of being generated from the topic about
supervision and regulation. By reading the speech we can verify this,

\begin{displayquote}[Ms Susan Schmidt Bies, Member of the Board of Governors of the US Federal
    Reserve System, at the Global Association of Risk Professionals' Basel II and Bakning
    Regulation Reform, Barcelona, 16 May 2006.]
In my remarks, I will focus primarily on the
choices and challenges associated with Basel II implementation. In particular, I want to reaffirm the
Federal Reserve’s commitment to Basel II and the need for continual evolution in risk measurement
and management at our largest banks and then discuss a few key aspects of Basel II implementation
in the United States. Given the international audience here today, I also plan to offer some thoughts
on cross-border implementation issues associated with Basel II, including so-called home-host issues.
\end{displayquote}

Overall in this paper, the manual validation (of a small subset) of the speeches suggests that the
dimension reduction to topic space captures the content of the documents well.

\subsection{Persistence in topics\label{section:results_persistence}}
\afterpage{
\begin{landscape}\centering\vspace*{\fill}
\begin{table}[h]
    \centering
    \scalebox{0.65}{
    \input{./ar_q_mod}
}
    \caption{Estimated coefficients from the AR(1) models,
    using quarterly data. Each equation is represented as a column.
    The t-statistics (reported in parenthesis) are based on
    \textcite{newey1987} standard errors with 1 lag.}
    \label{table:ar_q}

    \centering
    \scalebox{0.65}{
    \input{./ar_m_mod}
}
    \caption{Estimated coefficients from the AR(1) models,
    using monthly data. Each equation is represented as a column.
    The t-statistics (reported in parenthesis) are based on
    \textcite{newey1987} standard errors with 1 lag.}
    \label{table:ar_m}
\end{table}
\vfill
\end{landscape}
}

Next we construct a set of autoregressive (AR) models to investigate the persistence in the
estimated global topics.
The analyses are based on quarterly data following \textcite{armelius2020} and \textcite{nyman2018}, as
well as monthly data for robustness. The following AR(1) model is estimated for each global topic,

\begin{equation}\label{eq:ar}
    \theta_{k, t} = \alpha_{k} + \varphi_{k} \theta_{k, t-1} + \bm{\beta}_{k} \bm{X}_{t} +
    \bm{\gamma}_{k} \bm{X}_{t-1} + \epsilon_{k, t}.
\end{equation}

Here $\theta_{k, t}$ is the average probability for the classified documents at time $t$ for global
topic $k$, where $k = 1, \dots, K$ and  $t=1, \dots, T$. $\bm{X_}{t}$
is a vector of region specific control variables at time $t$, used in the Fed case study
in Section \ref{section:fed}.

Tables \ref{table:ar_q} and \ref{table:ar_m} report the estimated coefficients from the AR(1) model
in Equation \ref{eq:ar} without regional control variables,
using aggregated quarterly and monthly observations for each topic respectively.
The tables show strong significant autoregressive effects in the majority of the
topics.\footnote{The results are robust to estimation in a VAR system, where the
autoregressive effects dominate the  cross-sectional effects.} The
results indicate large persistence in the topics, on both
quarterly and monthly basis. When central
banks talk about a given global topic they tend to continue to do so for some time.
An explanation is that the underlying
macroeconomic variables that the topics reflect are themselves persistent, which we will
control for in the second part of the analysis in Section \ref{section:fed}. Another explanation is
that narratives drive the persistence in the topics. These narratives can either be part of
the global economy or narratives set by the central banks themselves.
A few topics are not showing any autoregressive effects.
The topic about supervision and regulation does not exhibit any significant persistence, indicating
that the topic is talked about sporadically throughout the sample, without specific trends.

%% file: topic_0_mod.tex
\scalebox{0.55}{
\begin{tabular}{a|b|a|b|a|b|a}\hline\hline
1997&1998&1999&...&2018&2019&2020\\
\hline
 Supervisor (0.024) & Supervisor (0.024) & Supervisor (0.022) &...& Regulation (0.016) & Regulation (0.016) & Regulation (0.016) \\
 Standard (0.021) & Standard (0.02) & Standard (0.017) &...& Requirement (0.009) & Stress Test (0.01) & Stress Test (0.01) \\
 Approach (0.02) & Approach (0.018) & Approach (0.016) &...& Stress Test (0.009) & Approach (0.009) & Approach (0.01) \\
 Supervisory (0.009) & Supervisory (0.009) & Supervisory (0.008) &...& Approach (0.008) & Requirement (0.009) & Requirement (0.009) \\
 Internal (0.008) & Internal (0.008) & Market Discipline (0.008) &...& Capital Requirement (0.008) & Capital Requirement (0.008) & Rule (0.008) \\
 Institution (0.008) & Institution (0.008) & Institution (0.007) &...& Rule (0.007) & Rule (0.007) & Capital Requirement (0.008) \\
 Risk Management (0.007) & Market Discipline (0.008) & Internal (0.007) &...& Regulatory (0.007) & Regulatory (0.007) & Regulatory (0.008) \\
 Market Discipline (0.007) & Risk Management (0.007) & Risk Management (0.007) &...& Regime (0.006) & Framework (0.007) & Framework (0.007) \\
 Capital Requirement (0.006) & Exposure (0.006) & Exposure (0.007) &...& Standard (0.006) & Regime (0.006) & Regime (0.006) \\
 Exposure (0.006) & Proposal (0.006) & Proposal (0.006) &...& Framework (0.005) & Stress Testing (0.006) & Stress Testing (0.006) \\
\hline\hline
\end{tabular}
}

%% file: ar_q_mod.tex
{
\def\sym#1{\ifmmode^{#1}\else\(^{#1}\)\fi}
\begin{tabular}{b a b a b a b a b a b a b a b a b a b a b a b a b a b a b a}
\hline\hline
            &\multicolumn{1}{c}{(1)}&\multicolumn{1}{c}{(2)}&\multicolumn{1}{c}{(3)}&\multicolumn{1}{c}{(4)}&\multicolumn{1}{c}{(5)}&\multicolumn{1}{c}{(6)}&\multicolumn{1}{c}{(7)}&\multicolumn{1}{c}{(8)}&\multicolumn{1}{c}{(9)}&\multicolumn{1}{c}{(10)}&\multicolumn{1}{c}{(11)}&\multicolumn{1}{c}{(12)}&\multicolumn{1}{c}{(13)}&\multicolumn{1}{c}{(14)}&\multicolumn{1}{c}{(15)}&\multicolumn{1}{c}{(16)}&\multicolumn{1}{c}{(17)}\\
            &\multicolumn{1}{c}{\shortstack{Regu-\\lation}}&\multicolumn{1}{c}{Money}&\multicolumn{1}{c}{\shortstack{Financial\\stability}}&\multicolumn{1}{c}{Payment}&\multicolumn{1}{c}{\shortstack{Euro\\area}}&\multicolumn{1}{c}{\shortstack{Comm-\\unity}}&\multicolumn{1}{c}{\shortstack{Financial\\system}}&\multicolumn{1}{c}{\shortstack{Unemp-\\loyment}}&\multicolumn{1}{c}{Investor}&\multicolumn{1}{c}{\shortstack{Inno-\\vation}}&\multicolumn{1}{c}{\shortstack{Liqui-\\dity}}&\multicolumn{1}{c}{Model}&\multicolumn{1}{c}{Fiscal}&\multicolumn{1}{c}{\shortstack{Produ-\\ctivity}}&\multicolumn{1}{c}{Trade}&\multicolumn{1}{c}{FOMC}&\multicolumn{1}{c}{Deflation}\\
\hline
1 quarter lag       &       0.228         &       0.385\sym{***}&       0.468\sym{***}&       0.587\sym{***}&       0.892\sym{***}&       0.644\sym{***}&       0.767\sym{***}&       0.719\sym{***}&       0.693\sym{***}&       0.525\sym{***}&       0.770\sym{***}&       0.180         &       0.640\sym{***}&       0.579\sym{***}&       0.638\sym{***}&       0.380\sym{**} &       0.926\sym{***}\\
    (No controls)            &      (1.85)         &      (4.77)         &      (4.21)         &      (7.33)         &     (15.51)         &      (5.47)         &     (10.25)         &      (9.47)         &      (6.58)         &      (6.25)         &     (15.56)         &      (1.76)         &      (7.03)         &      (6.28)         &      (7.16)         &      (3.33)         &     (23.67)         \\
[1em]
Constants   &      0.0230\sym{***}&     0.00756\sym{***}&      0.0188\sym{***}&      0.0109\sym{***}&     0.00387\sym{*}  &     0.00985\sym{***}&      0.0113\sym{**} &     0.00964\sym{**} &     0.00626\sym{***}&      0.0125\sym{***}&     0.00789\sym{***}&      0.0387\sym{***}&      0.0126\sym{***}&      0.0175\sym{***}&      0.0133\sym{***}&      0.0206\sym{***}&     0.00234\sym{*}  \\
            &      (5.85)         &      (6.06)         &      (5.23)         &      (4.99)         &      (2.59)         &      (3.56)         &      (3.30)         &      (3.40)         &      (3.44)         &      (5.47)         &      (3.54)         &      (7.07)         &      (3.76)         &      (4.48)         &      (3.85)         &      (5.59)         &      (2.50)         \\
\hline
\(N\)       &          95         &          95         &          95         &          95         &          95         &          95         &          95         &          95         &          95         &          95         &          95         &          95         &          95         &          95         &          95         &          95         &          95         \\
\hline\hline
\multicolumn{18}{l}{\footnotesize \textit{t} statistics in parentheses}\\
\multicolumn{18}{l}{\footnotesize \sym{*} \(p<0.05\), \sym{**} \(p<0.01\), \sym{***} \(p<0.001\)}\\
\end{tabular}
}

%% file: ar_m_mod.tex
{
\def\sym#1{\ifmmode^{#1}\else\(^{#1}\)\fi}
\begin{tabular}{b a b a b a b a b a b a b a b a b a b a b a b a b a b a b a}
\hline\hline
            &\multicolumn{1}{c}{(1)}&\multicolumn{1}{c}{(2)}&\multicolumn{1}{c}{(3)}&\multicolumn{1}{c}{(4)}&\multicolumn{1}{c}{(5)}&\multicolumn{1}{c}{(6)}&\multicolumn{1}{c}{(7)}&\multicolumn{1}{c}{(8)}&\multicolumn{1}{c}{(9)}&\multicolumn{1}{c}{(10)}&\multicolumn{1}{c}{(11)}&\multicolumn{1}{c}{(12)}&\multicolumn{1}{c}{(13)}&\multicolumn{1}{c}{(14)}&\multicolumn{1}{c}{(15)}&\multicolumn{1}{c}{(16)}&\multicolumn{1}{c}{(17)}\\
            &\multicolumn{1}{c}{\shortstack{Regu-\\lation}}&\multicolumn{1}{c}{Money}&\multicolumn{1}{c}{\shortstack{Financial\\stability}}&\multicolumn{1}{c}{Payment}&\multicolumn{1}{c}{\shortstack{Euro\\area}}&\multicolumn{1}{c}{\shortstack{Comm-\\unity}}&\multicolumn{1}{c}{\shortstack{Financial\\system}}&\multicolumn{1}{c}{\shortstack{Unemp-\\loyment}}&\multicolumn{1}{c}{Investor}&\multicolumn{1}{c}{\shortstack{Inno-\\vation}}&\multicolumn{1}{c}{\shortstack{Liqui-\\dity}}&\multicolumn{1}{c}{Model}&\multicolumn{1}{c}{Fiscal}&\multicolumn{1}{c}{\shortstack{Produ-\\ctivity}}&\multicolumn{1}{c}{Trade}&\multicolumn{1}{c}{FOMC}&\multicolumn{1}{c}{Deflation}\\
\hline
1 month lag       &       0.118         &       0.222\sym{**} &       0.286\sym{***}&       0.363\sym{***}&       0.716\sym{***}&       0.258\sym{***}&       0.473\sym{***}&       0.426\sym{***}&       0.274\sym{***}&       0.181\sym{**} &       0.731\sym{***}&       0.140\sym{*}  &       0.430\sym{***}&       0.380\sym{***}&       0.417\sym{***}&       0.131         &       0.669\sym{***}\\
    (No controls)            &      (1.82)         &      (2.69)         &      (3.78)         &      (5.27)         &     (12.62)         &      (3.88)         &      (7.65)         &      (7.75)         &      (3.57)         &      (3.26)         &     (14.22)         &      (2.09)         &      (6.41)         &      (6.16)         &      (5.94)         &      (1.90)         &     (12.26)         \\
[1em]
Constants   &      0.0249\sym{***}&     0.00962\sym{***}&      0.0247\sym{***}&      0.0165\sym{***}&     0.00923\sym{***}&      0.0191\sym{***}&      0.0257\sym{***}&      0.0209\sym{***}&      0.0139\sym{***}&      0.0210\sym{***}&     0.00842\sym{***}&      0.0397\sym{***}&      0.0201\sym{***}&      0.0265\sym{***}&      0.0214\sym{***}&      0.0288\sym{***}&     0.00835\sym{***}\\
            &     (10.13)         &      (9.06)         &      (9.46)         &      (8.51)         &      (5.06)         &      (8.08)         &      (8.16)         &      (8.59)         &      (8.29)         &     (12.00)         &      (5.93)         &     (11.88)         &      (7.51)         &      (8.64)         &      (8.20)         &     (10.44)         &      (5.59)         \\
\hline
\(N\)       &         287         &         287         &         287         &         287         &         287         &         287         &         287         &         287         &         287         &         287         &         287         &         287         &         287         &         287         &         287         &         287         &         287         \\
\hline\hline
\multicolumn{18}{l}{\footnotesize \textit{t} statistics in parentheses}\\
\multicolumn{18}{l}{\footnotesize \sym{*} \(p<0.05\), \sym{**} \(p<0.01\), \sym{***} \(p<0.001\)}\\
\end{tabular}
}

%% file: fed.tex
In this section, Equation \ref{eq:ar} from Section \ref{section:results_persistence} is estimated
using the Fed speech data alone, together with regional financial control variables.
Central banks tend to address the current economic environment in their communication. Therefore, underlying
financial and macroeconomic variables play an important role in determining the topics of central bank
communication. Central banks display heterogeneity in the topics addressed, as well as which variables affect
their communication. The Fed is more likely to talk about topics related to the US inflation and
the US stock market than topics related to European macroeconomic and financial conditions.
By selecting speeches from one central bank alone, in this case the Fed, it is possible
to control for the regional variables associated with that bank's speeches.

\begin{figure}[h]
    \includegraphics[width=1\textwidth]{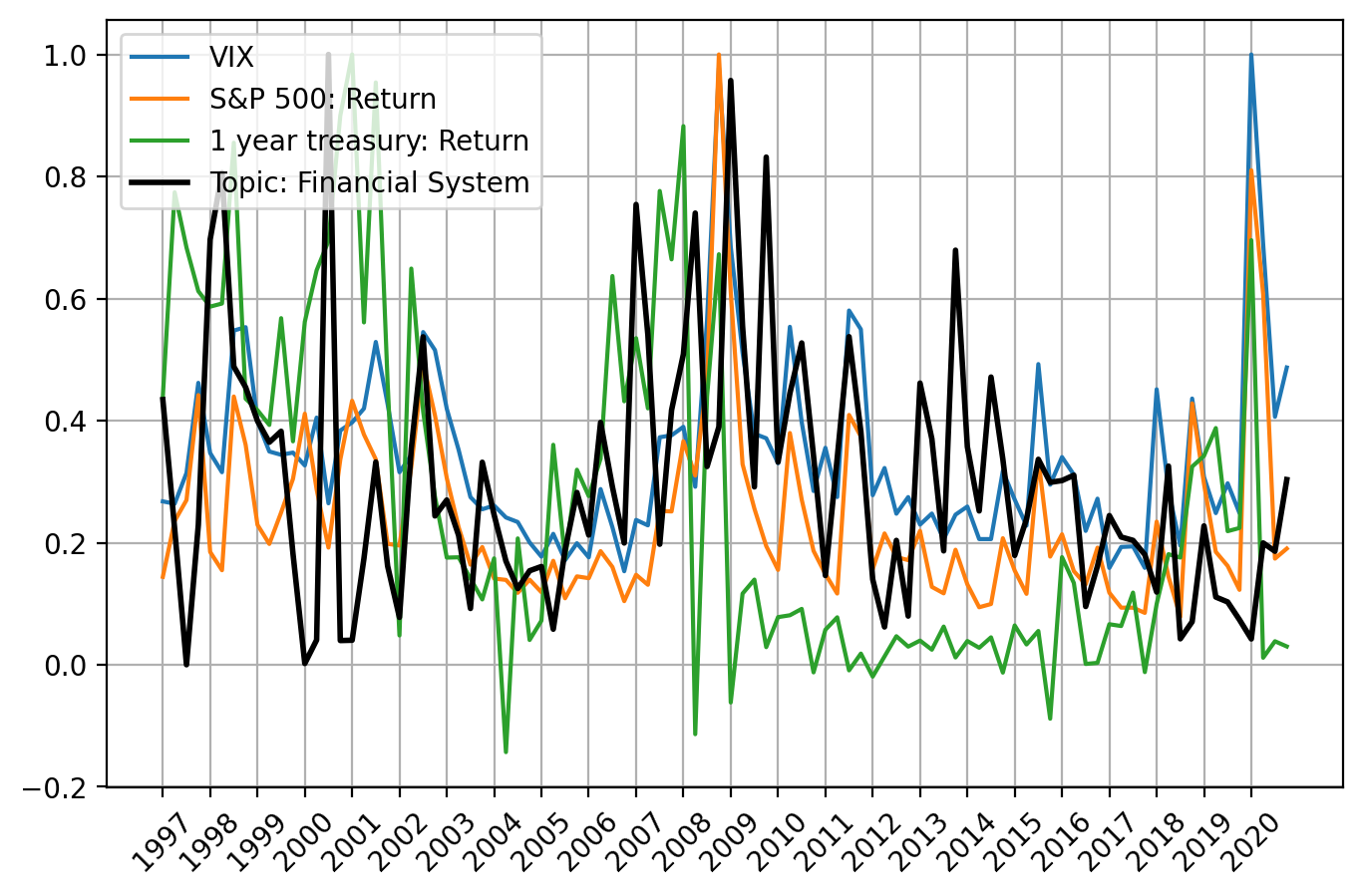}
    \centering
    \caption{Plot of average quarterly normalized probability of the topic about the financial
    system, together with three normalized control variables.}
    \label{fig:controls}
\end{figure}

Figure \ref{fig:controls} illustrates the co-movements between the probability of the topic about the
financial system in speeches from the Fed, together with the CBOE Volatility Index (VIX),
the S\&P 500 Index returns, and 1 year US treasury returns. The figure suggests that there might be a
relationship between these variables. Thus, by controlling for the financial variables discussed in
the topic about the financial system we might be able to explain the autoregressive feature of the
topic. If the persistence is fully explained by the controls, it would suggest that no
other effects, such as global narratives or local central bank narratives, are driving the previous
observed persistence.

Table \ref{table:fed_q} reports the AR(1) coefficients of the average topic probabilities from 1,841 classified
speeches from the Fed and the New York Fed, from 1997 through 2020, without any control variables.
Quarterly data are used alone
as there are months during the sample period in which no speeches by the Fed were held.
The speeches are classified using the DTM model in Section \ref{section:results}. The model is not
re-estimated but
trained on the full corpus, including speeches from the other 8 central banks analyzed in this paper.
Hence, the model is less descriptive of the Fed data, which might be a reason
why the results are weaker compared to the previous results in Table \ref{table:ar_q}, Section
\ref{section:results}.\footnote{The AR(1) results are stronger when estimating using topic
probabilities from  a DTM model trained on the Fed corpus alone, see table \ref{table:ar_control}
in Appendix \ref{section:A_fed}.}

The autoregressive properties in the global topics are weakened by adding the controls,
but not fully explained, as seen in Table \ref{table:fed_q_c}.
Compared to the estimation without the control variables the results show smaller coefficients and lower
significance of the persistent topics.
But, most topics that exhibit persistence without the controls are also doing so with the
controls. Two exceptions are the topic about the financial system and the topic about trade, where
the persistence seems to be fully explained by the control variables.

The results from the Fed case study suggest that there are factors other than the underlying
financial variables driving the topics' persistence. Further, the results are consistent with the
theory of narrative economics and propose that the communication on topic level is
\textit{story-based}.
Story-based communication is more easily spread in conversations, news and
social media \parencite{shiller2017}, and coming from a central bank it is more accessible for the general public
\parencite{haldane2018}.

\afterpage{
\begin{landscape}\centering\vspace*{\fill}
\begin{table}[h]
    \centering
    \scalebox{0.65}{
    \input{./fed_q_mod}
}
    \caption{Estimated coefficients from the AR(1) models, using quarterly data from the Fed. Each
    equation is represented as a column. The t-statistics (reported in parenthesis) are based on
    \textcite{newey1987} standard errors with 1 lag.}
    \label{table:fed_q}

    \centering
    \scalebox{0.65}{
    \input{./fed_q_ctrl_mod}
}
    \caption{Estimated coefficients from the AR(1) models with control variables,
    using quarterly data from the Fed. Each equation is represented as a column.
    The t-statistics (reported in parenthesis) are based on
    \textcite{newey1987} standard errors with 1 lag.}
    \label{table:fed_q_c}
\end{table}
\vfill
\end{landscape}
}

%% file: fed_q_mod.tex
{
\def\sym#1{\ifmmode^{#1}\else\(^{#1}\)\fi}
\begin{tabular}{b a b a b a b a b a b a b a b a b a b a b a b a b a b a b a}
\hline\hline
            &\multicolumn{1}{c}{(1)}&\multicolumn{1}{c}{(2)}&\multicolumn{1}{c}{(3)}&\multicolumn{1}{c}{(4)}&\multicolumn{1}{c}{(5)}&\multicolumn{1}{c}{(6)}&\multicolumn{1}{c}{(7)}&\multicolumn{1}{c}{(8)}&\multicolumn{1}{c}{(9)}&\multicolumn{1}{c}{(10)}&\multicolumn{1}{c}{(11)}&\multicolumn{1}{c}{(12)}&\multicolumn{1}{c}{(13)}&\multicolumn{1}{c}{(14)}&\multicolumn{1}{c}{(15)}&\multicolumn{1}{c}{(16)}&\multicolumn{1}{c}{(17)}\\
            &\multicolumn{1}{c}{\shortstack{Regu-\\lation}}&\multicolumn{1}{c}{Money}&\multicolumn{1}{c}{\shortstack{Financial\\stability}}&\multicolumn{1}{c}{Payment}&\multicolumn{1}{c}{\shortstack{Euro\\area}}&\multicolumn{1}{c}{\shortstack{Comm-\\unity}}&\multicolumn{1}{c}{\shortstack{Financial\\system}}&\multicolumn{1}{c}{\shortstack{Unemp-\\loyment}}&\multicolumn{1}{c}{Investor}&\multicolumn{1}{c}{\shortstack{Inno-\\vation}}&\multicolumn{1}{c}{\shortstack{Liqui-\\dity}}&\multicolumn{1}{c}{Model}&\multicolumn{1}{c}{Fiscal}&\multicolumn{1}{c}{\shortstack{Produ-\\ctivity}}&\multicolumn{1}{c}{Trade}&\multicolumn{1}{c}{FOMC}&\multicolumn{1}{c}{Deflation}\\
\hline
    1 quater lag      &       0.121         &       0.163         &       0.236         &       0.374\sym{**} &       0.685\sym{***}&       0.368\sym{*}  &       0.280\sym{*}  &       0.430\sym{***}&       0.400\sym{***}&       0.151         &       0.690\sym{***}&      0.0870         &       0.468\sym{***}&       0.478\sym{***}&       0.342\sym{***}&       0.139         &       0.564\sym{***}\\
    (No controls)        &      (1.04)         &      (1.44)         &      (1.91)         &      (3.07)         &      (7.68)         &      (2.26)         &      (2.25)         &      (5.71)         &      (4.65)         &      (1.40)         &      (6.88)         &      (0.73)         &      (4.11)         &      (4.40)         &      (4.38)         &      (1.73)         &      (5.18)         \\
[1em]
Constants   &      0.0267\sym{***}&      0.0100\sym{***}&      0.0297\sym{***}&      0.0174\sym{***}&      0.0104\sym{***}&      0.0212\sym{***}&      0.0353\sym{***}&      0.0188\sym{***}&      0.0120\sym{***}&      0.0246\sym{***}&      0.0105\sym{***}&      0.0435\sym{***}&      0.0192\sym{***}&      0.0216\sym{***}&      0.0255\sym{***}&      0.0315\sym{***}&      0.0117\sym{***}\\
            &      (6.35)         &      (6.36)         &      (5.96)         &      (4.61)         &      (3.71)         &      (4.34)         &      (4.97)         &      (6.07)         &      (4.60)         &      (6.31)         &      (3.90)         &      (6.92)         &      (3.95)         &      (4.92)         &      (6.14)         &      (7.09)         &      (4.09)         \\
\hline
\(N\)       &          95         &          95         &          95         &          95         &          95         &          95         &          95         &          95         &          95         &          95         &          95         &          95         &          95         &          95         &          95         &          95         &          95         \\
\hline\hline
\multicolumn{18}{l}{\footnotesize \textit{t} statistics in parentheses}\\
\multicolumn{18}{l}{\footnotesize \sym{*} \(p<0.05\), \sym{**} \(p<0.01\), \sym{***} \(p<0.001\)}\\
\end{tabular}
}

%% file: fed_q_ctrl_mod.tex
{
\def\sym#1{\ifmmode^{#1}\else\(^{#1}\)\fi}
\begin{tabular}{b a b a b a b a b a b a b a b a b a b a b a b a b a b a b a}
\hline\hline
            &\multicolumn{1}{c}{(1)}&\multicolumn{1}{c}{(2)}&\multicolumn{1}{c}{(3)}&\multicolumn{1}{c}{(4)}&\multicolumn{1}{c}{(5)}&\multicolumn{1}{c}{(6)}&\multicolumn{1}{c}{(7)}&\multicolumn{1}{c}{(8)}&\multicolumn{1}{c}{(9)}&\multicolumn{1}{c}{(10)}&\multicolumn{1}{c}{(11)}&\multicolumn{1}{c}{(12)}&\multicolumn{1}{c}{(13)}&\multicolumn{1}{c}{(14)}&\multicolumn{1}{c}{(15)}&\multicolumn{1}{c}{(16)}&\multicolumn{1}{c}{(17)}\\
            &\multicolumn{1}{c}{\shortstack{Regu-\\lation}}&\multicolumn{1}{c}{Money}&\multicolumn{1}{c}{\shortstack{Financial\\stability}}&\multicolumn{1}{c}{Payment}&\multicolumn{1}{c}{\shortstack{Euro\\area}}&\multicolumn{1}{c}{\shortstack{Comm-\\unity}}&\multicolumn{1}{c}{\shortstack{Financial\\system}}&\multicolumn{1}{c}{\shortstack{Unemp-\\loyment}}&\multicolumn{1}{c}{Investor}&\multicolumn{1}{c}{\shortstack{Inno-\\vation}}&\multicolumn{1}{c}{\shortstack{Liqui-\\dity}}&\multicolumn{1}{c}{Model}&\multicolumn{1}{c}{Fiscal}&\multicolumn{1}{c}{\shortstack{Produ-\\ctivity}}&\multicolumn{1}{c}{Trade}&\multicolumn{1}{c}{FOMC}&\multicolumn{1}{c}{Deflation}\\
\hline
1 quarter lag       &      0.0211         &       0.170         &       0.174         &       0.381\sym{**} &       0.583\sym{***}&       0.309         &       0.262         &       0.270\sym{*}  &       0.367\sym{***}&      0.0928         &       0.542\sym{***}&     -0.0121         &       0.334\sym{*}  &       0.421\sym{***}&       0.127         &       0.114         &       0.479\sym{***}\\
    (With controls)            &      (0.18)         &      (1.37)         &      (1.60)         &      (2.81)         &      (5.48)         &      (1.93)         &      (1.88)         &      (2.15)         &      (3.79)         &      (0.79)         &      (4.51)         &     (-0.10)         &      (2.25)         &      (4.06)         &      (0.98)         &      (1.31)         &      (3.94)         \\
[1em]
Constants   &      0.0497\sym{***}&      0.0112\sym{**} &      0.0258         &      0.0191\sym{*}  &      0.0332\sym{*}  &      0.0122         &      0.0208         &      0.0192\sym{**} &      0.0201\sym{*}  &      0.0401\sym{***}&     -0.0118         &      0.0572\sym{***}&      0.0183         &      0.0405\sym{**} &      0.0310\sym{**} &      0.0472\sym{***}&      0.0206         \\
            &      (4.49)         &      (2.77)         &      (1.58)         &      (2.01)         &      (2.35)         &      (0.83)         &      (1.82)         &      (2.85)         &      (2.46)         &      (3.95)         &     (-0.74)         &      (4.37)         &      (1.43)         &      (3.36)         &      (3.03)         &      (3.60)         &      (1.84)         \\
\hline
\(N\)       &          95         &          95         &          95         &          95         &          95         &          95         &          95         &          95         &          95         &          95         &          95         &          95         &          95         &          95         &          95         &          95         &          95         \\
\hline\hline
\multicolumn{18}{l}{\footnotesize \textit{t} statistics in parentheses}\\
\multicolumn{18}{l}{\footnotesize \sym{*} \(p<0.05\), \sym{**} \(p<0.01\), \sym{***} \(p<0.001\)}\\
\end{tabular}
}

%% file: conclusion.tex
The empirical findings show that central banks talk about a wide range of global topics, not all
immediately related to the traditional theory of central bank communication. The topics are consistent
with the literature arguing that the communication is not directly targeting the general public.
However, with a broader set of topics, central banks can
reveal private information and prepare society for long term monetary policy shifts and structural changes.
Topic trends occur and vocabulary changes over time, but most topics have
significant probability mass throughout the sample period, even at times when the interest rate is
close to the ELB.

Furthermore, the topics are well captured by Dynamic Topic Models. Both in terms of
quantitative measures, such as coherence scores, as well as manual
investigation linking the topics to the representative documents. Thus, the dimension
reduction of the corpus to topic space is able to, in a meaningful way, capture the relevant
central bank communication. This encourages the use of topic modelling, and more specifically DTM,
in other social science applications with similar data.

Topic modelling has an interesting application in estimating narratives.
The observed topic persistence is consistent with the theory of narrative economics and proposes that the
central bank communication on topic level
is story-based. The evolution of word-probabilities within the topics are also consistent with the epidemiology
models of narrative economics.

%% file: appendix.tex
\section{Results}\label{section:A_results}

Table \ref{table:A_topics} shows the full set of estimated topics.
Table \ref{table:A_summary} shows the number of speeches per central bank per year.
Table \ref{table:A_averages} shows the average topic distribution for each central bank. The average
distributions are calculated by taking the mean over the topic distributions for all classified
documents for each central bank.
Table \ref{table:fed_q_c_full} shows the full set of coefficients from the Fed caste study
in Section \ref{section:fed}, including coefficients for the control variables.

\begin{table}[h]
    \centering
    \input{./topics_table_mod}
    \caption{Estimated topics, both local and global, and respectively probabilities in corpus.}
    \label{table:A_topics}
\end{table}

\begin{table}[h]
    \centering
    \input{./data_summary_table_mod}
    \caption{Number of speeches per central bank per year.}
    \label{table:A_summary}
\end{table}

\afterpage{
\begin{landscape}\centering\vspace*{\fill}
\begin{table}[h]
    \centering
    \scalebox{0.9}{
    \input{./averages_table_mod}
}
    \caption{Central banks' average topic distribution.}
    \label{table:A_averages}
\end{table}
\vfill
\end{landscape}
}

\begin{table}[h]
    \centering
    \scalebox{0.46}{
    \input{./fed_q_ctrl_mod_full}
}
    \caption{Estimated coefficients from the AR(1) models with control variables,
    using quarterly data from the Fed. Each equation is represented as a column.
    The t-statistics (reported in parenthesis) are based on
    \textcite{newey1987} standard errors with 1 lag.}
    \label{table:fed_q_c_full}
\end{table}

\section{Fed robustness}\label{section:A_fed}
\begin{figure}[h]
    \includegraphics[width=1\textwidth]{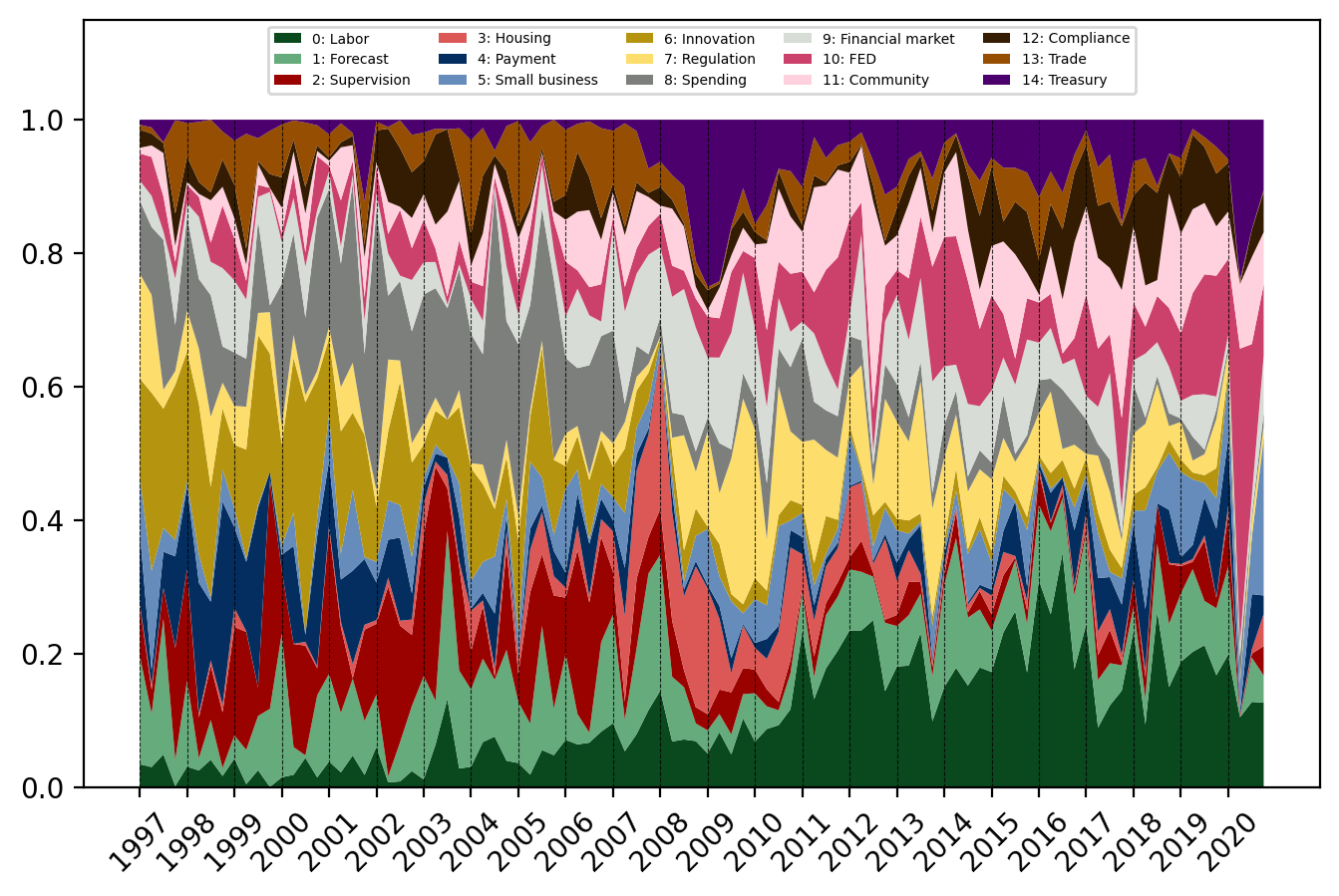}
    \centering
    \caption{Plot of the average probabilities of topics over each quarter.}
    \label{fig:fed_stackplot}
\end{figure}

Figure \ref{fig:fed_stackplot} shows the average normalized probability distributions for the topics
for each quarter, given by the classified documents using DTM trained on the data from
the Fed and the New York Fed alone. In
each quarter all speeches are classified, their probability distributions averaged, and plotted.

Compared to the main model in the paper, one can see that the topic about supervision and regulation has split into
two topics. The patterns of the topics are familiar, and here expressed by a higher probability of
the topic related to supervision in the beginning of the corpus, and a higher probability of the
topic related to regulation in the end.

Table \ref{table:ar_control} reports the estimated coefficients of the AR(1) models estimated using
data from a topic model trained on data from the Fed and the New York Fed alone.

\begin{table}[h]
    \centering
    \scalebox{0.5}{
    \input{./ar_q_control_mod}
}
    \caption{Estimated coefficients from the AR(1) models, using quarterly data from the Fed. The
    documents are classified with a DTM model trained on the Fed speeches alone. Each
    equation is represented as a column. The t-statistics (reported in parenthesis) are based on
    \textcite{newey1987} standard errors with 1 lag.}
    \label{table:ar_control}
\end{table}

%% file: topics_table_mod.tex
\scalebox{0.7}{
\begin{tabular}{a|b}\hline\hline
Topics&Average $\mathrm{P}$ ($\%$) in corpus\\
\hline
0: Supervision&3.17\\
1: Money&1.61\\
2: Financial stability&7.25\\
3: Payment&2.77\\
4: Euro area&2.63\\
5: Community&3.3\\
6: Financial system&2.56\\
7: Unemployment&1.61\\
8: Investor&5.33\\
9: Innovation&2.63\\
10: Liquidity&1.93\\
11: Model&4.37\\
12: Fiscal&5.51\\
13: Productivity&3.88\\
14: Trade&4.37\\
15: FOMC&2.14\\
16: Deflation&2.55\\
\hline
Sum global: & 57.61\\
\hline
17: Australia&2.97\\
18: Canada&7.02\\
19: ECB&3.01\\
20: Euro area&4.54\\
21: Federal reserve&4.67\\
22: Japan economy&1.76\\
23: Japan&5.29\\
24: Norway&2.12\\
25: Riksbank&2.89\\
26: Residual&4.08\\
27: SNB&2.54\\
28: UK&1.53\\
\hline
Sum local: &42.39\\
\hline
Mean&3.45\\
\hline
Sum&100.0\\
\hline\hline
\end{tabular}
}

%% file: data_summary_table_mod.tex
\scalebox{0.5}{
\begin{tabular}{a|b|a|b|a|b|a|b|a|b|a|b|a|b|a|b|a|b|a|b|a|b|a|b|a|b|a|b}\hline\hline
&1997&1998&1999&2000&2001&2002&2003&2004&2005&2006&2007&2008&2009&2010&2011&2012&2013&2014&2015&2016&2017&2018&2019&2020&Sum\\
\hline
CAN&7&6&9&9&12&18&18&18&24&25&29&24&24&23&31&25&28&22&23&27&26&29&19&23&499\\
ENG&13&15&12&22&21&14&7&10&12&17&17&21&28&38&34&25&36&37&47&38&37&44&48&32&625\\
JPN&21&16&15&22&16&21&28&15&20&14&19&18&32&32&45&50&42&44&35&33&39&27&40&24&668\\
FED&52&43&42&39&48&55&55&97&86&76&85&88&70&73&62&51&59&46&57&42&54&49&79&57&1465\\
NOR&0&0&6&7&10&19&14&13&20&20&15&16&17&16&11&10&10&8&8&8&10&9&10&5&262\\
ECB&0&8&46&45&36&40&31&56&51&58&88&149&130&121&139&103&152&130&136&120&169&135&143&99&2185\\
NYC&5&7&5&7&2&4&3&8&10&12&9&4&11&23&22&18&29&26&34&32&29&26&28&22&376\\
AUS&9&12&9&12&8&12&11&9&13&9&10&17&17&29&32&27&26&26&34&25&33&38&34&23&475\\
SWE&11&13&28&31&29&29&21&24&23&36&22&27&23&25&28&15&17&13&9&11&5&12&8&6&466\\
CHE&1&1&3&2&4&11&11&28&27&27&26&29&25&19&16&21&15&14&16&15&10&16&14&7&358\\
\hline
Sum&119&121&175&196&186&223&199&278&286&294&320&393&377&399&420&345&414&366&399&351&412&385&423&298&7379\\
\hline\hline
\end{tabular}
}

%% file: averages_table_mod.tex
\scalebox{0.7}{
\begin{tabular}{a|b|a|b|a|b|a|b|a|b|a|b|a|b|a|b|a|b|a|b|a|b|a|b|a|b|a|b|a|b}\hline\hline
&0&1&2&3&4&5&6&7&8&9&10&11&12&13&14&15&16&17&18&19&20&21&22&23&24&25&26&27&28\\
\hline
CAN&0.009&0.012&0.501&0.018&0.011&0.011&0.002&0.01&0.001&0.002&0.003&0.015&0.046&0.001&0.019&0.016&0.039&0.021&0.067&0.004&0.002&0.064&0.007&0.002&0.004&0.035&0.064&0.007&0.008\\
ENG&0.04&0.016&0.042&0.035&0.03&0.015&0.231&0.013&0.001&0.0&0.003&0.027&0.082&0.001&0.055&0.052&0.035&0.026&0.153&0.02&0.001&0.037&0.006&0.001&0.002&0.017&0.04&0.005&0.016\\
JPN&0.004&0.01&0.004&0.015&0.02&0.012&0.002&0.003&0.001&0.255&0.003&0.006&0.022&0.001&0.013&0.011&0.022&0.015&0.029&0.004&0.434&0.038&0.005&0.002&0.002&0.02&0.033&0.005&0.013\\
FED&0.086&0.008&0.039&0.01&0.029&0.109&0.001&0.003&0.0&0.001&0.109&0.027&0.085&0.0&0.017&0.017&0.031&0.04&0.055&0.002&0.002&0.046&0.004&0.001&0.001&0.099&0.036&0.132&0.008\\
NOR&0.009&0.022&0.008&0.018&0.033&0.011&0.004&0.009&0.522&0.001&0.003&0.017&0.029&0.002&0.084&0.012&0.026&0.017&0.029&0.006&0.001&0.061&0.01&0.018&0.004&0.01&0.018&0.004&0.011\\
ECB&0.01&0.016&0.005&0.082&0.026&0.006&0.002&0.087&0.001&0.0&0.001&0.009&0.023&0.002&0.014&0.009&0.016&0.026&0.042&0.192&0.002&0.057&0.095&0.002&0.179&0.018&0.023&0.002&0.055\\
NYC&0.103&0.006&0.04&0.021&0.023&0.129&0.001&0.012&0.001&0.001&0.064&0.018&0.16&0.001&0.01&0.026&0.019&0.064&0.072&0.006&0.004&0.036&0.004&0.001&0.001&0.034&0.052&0.086&0.007\\
AUS&0.009&0.013&0.071&0.014&0.031&0.011&0.007&0.006&0.001&0.001&0.002&0.274&0.039&0.001&0.068&0.043&0.021&0.037&0.171&0.001&0.002&0.035&0.008&0.003&0.001&0.025&0.089&0.007&0.009\\
SWE&0.023&0.027&0.003&0.033&0.024&0.004&0.002&0.008&0.003&0.001&0.002&0.03&0.026&0.001&0.114&0.007&0.019&0.018&0.039&0.02&0.002&0.042&0.014&0.489&0.002&0.017&0.018&0.003&0.007\\
CHE&0.025&0.034&0.013&0.031&0.037&0.022&0.003&0.01&0.002&0.001&0.002&0.015&0.039&0.378&0.042&0.022&0.027&0.033&0.046&0.046&0.004&0.051&0.023&0.01&0.016&0.012&0.035&0.003&0.017\\
\hline
Mean&0.032&0.016&0.073&0.028&0.026&0.033&0.026&0.016&0.053&0.026&0.019&0.044&0.055&0.039&0.044&0.022&0.025&0.03&0.07&0.03&0.045&0.047&0.018&0.053&0.021&0.029&0.041&0.025&0.015\\
\hline\hline
\end{tabular}
}

%% file: fed_q_ctrl_mod_full.tex
{
\def\sym#1{\ifmmode^{#1}\else\(^{#1}\)\fi}
\begin{tabular}{b a b a b a b a b a b a b a b a b a b a b a b a b a b a b a}
\hline\hline
            &\multicolumn{1}{c}{(1)}&\multicolumn{1}{c}{(2)}&\multicolumn{1}{c}{(3)}&\multicolumn{1}{c}{(4)}&\multicolumn{1}{c}{(5)}&\multicolumn{1}{c}{(6)}&\multicolumn{1}{c}{(7)}&\multicolumn{1}{c}{(8)}&\multicolumn{1}{c}{(9)}&\multicolumn{1}{c}{(10)}&\multicolumn{1}{c}{(11)}&\multicolumn{1}{c}{(12)}&\multicolumn{1}{c}{(13)}&\multicolumn{1}{c}{(14)}&\multicolumn{1}{c}{(15)}&\multicolumn{1}{c}{(16)}&\multicolumn{1}{c}{(17)}\\
            &\multicolumn{1}{c}{\shortstack{Regu-\\lation}}&\multicolumn{1}{c}{Money}&\multicolumn{1}{c}{\shortstack{Financial\\stability}}&\multicolumn{1}{c}{Payment}&\multicolumn{1}{c}{\shortstack{Euro\\area}}&\multicolumn{1}{c}{\shortstack{Comm-\\unity}}&\multicolumn{1}{c}{\shortstack{Financial\\system}}&\multicolumn{1}{c}{\shortstack{Unemp-\\loyment}}&\multicolumn{1}{c}{Investor}&\multicolumn{1}{c}{\shortstack{Inno-\\vation}}&\multicolumn{1}{c}{\shortstack{Liqui-\\dity}}&\multicolumn{1}{c}{Model}&\multicolumn{1}{c}{Fiscal}&\multicolumn{1}{c}{\shortstack{Produ-\\ctivity}}&\multicolumn{1}{c}{Trade}&\multicolumn{1}{c}{FOMC}&\multicolumn{1}{c}{Deflation}\\
\hline
1 Lag       &      0.0211         &       0.170         &       0.174         &       0.381\sym{**} &       0.583\sym{***}&       0.309         &       0.262         &       0.270\sym{*}  &       0.367\sym{***}&      0.0928         &       0.542\sym{***}&     -0.0121         &       0.334\sym{*}  &       0.421\sym{***}&       0.127         &       0.114         &       0.479\sym{***}\\
            &      (0.18)         &      (1.37)         &      (1.60)         &      (2.81)         &      (5.48)         &      (1.93)         &      (1.88)         &      (2.15)         &      (3.79)         &      (0.79)         &      (4.51)         &     (-0.10)         &      (2.25)         &      (4.06)         &      (0.98)         &      (1.31)         &      (3.94)         \\
[1em]
Inflation   &      -0.217         &      0.0326         &     -0.0803         &      -0.396         &      -0.305         &      -0.141         &       0.225         &      -0.145         &      -0.108         &      0.0291         &      -0.416         &       0.373         &       0.518         &       0.645         &       0.165         &       0.383         &       0.114         \\
            &     (-0.57)         &      (0.24)         &     (-0.13)         &     (-1.21)         &     (-1.08)         &     (-0.39)         &      (0.53)         &     (-0.46)         &     (-0.71)         &      (0.13)         &     (-1.20)         &      (1.31)         &      (1.15)         &      (1.41)         &      (0.44)         &      (1.37)         &      (0.35)         \\
[1em]
\shortstack{Lag\\Inflation}&      -0.165         &      -0.366\sym{**} &      -0.125         &       0.375         &      0.0950         &     -0.0224         &      -0.381         &       0.350         &      -0.157         &       0.204         &     -0.0432         &       0.212         &      0.0725         &      -0.175         &     -0.0773         &      -0.586         &      -0.525         \\
            &     (-0.49)         &     (-3.33)         &     (-0.30)         &      (1.09)         &      (0.24)         &     (-0.06)         &     (-0.84)         &      (1.32)         &     (-0.88)         &      (0.76)         &     (-0.11)         &      (0.69)         &      (0.17)         &     (-0.42)         &     (-0.26)         &     (-1.56)         &     (-1.84)         \\
[1em]
\shortstack{1 year\\bond}&      -0.793         &     -0.0410         &      -0.778         &      -0.251         &      -0.301         &      -0.925         &      -0.596         &       1.587\sym{**} &       0.237         &      -0.223         &      -1.578\sym{*}  &      -0.384         &      -0.684         &     -0.0599         &       2.037\sym{**} &      -0.259         &      0.0466         \\
            &     (-1.66)         &     (-0.14)         &     (-1.10)         &     (-0.39)         &     (-0.63)         &     (-1.59)         &     (-0.70)         &      (2.90)         &      (0.84)         &     (-0.49)         &     (-2.16)         &     (-0.70)         &     (-1.24)         &     (-0.06)         &      (3.12)         &     (-0.42)         &      (0.11)         \\
[1em]
\shortstack{Lag\\1 year\\bond}&    -0.00155         &       0.142         &      -0.323         &       0.493         &      -1.148\sym{*}  &     -0.0752         &       1.010         &      -0.161         &      -0.388         &       0.384         &       0.943         &       0.549         &      -0.953         &       0.611         &       0.151         &      -0.336         &      -1.151\sym{*}  \\
            &     (-0.00)         &      (0.41)         &     (-0.45)         &      (0.69)         &     (-2.15)         &     (-0.11)         &      (1.30)         &     (-0.28)         &     (-1.44)         &      (0.86)         &      (1.32)         &      (1.17)         &     (-1.37)         &      (0.82)         &      (0.25)         &     (-0.57)         &     (-2.33)         \\
[1em]
\shortstack{SP500\\returns}&      -0.244         &      -0.394\sym{*}  &       1.201\sym{*}  &      -0.275         &       0.424         &      0.0697         &      -0.299         &      0.0832         &      -0.635\sym{*}  &       0.166         &       0.313         &       1.319\sym{**} &     -0.0479         &       0.773         &     -0.0949         &       0.158         &      -0.213         \\
            &     (-0.47)         &     (-2.04)         &      (1.99)         &     (-0.47)         &      (1.13)         &      (0.11)         &     (-0.48)         &      (0.19)         &     (-2.49)         &      (0.50)         &      (0.63)         &      (2.93)         &     (-0.10)         &      (1.35)         &     (-0.22)         &      (0.38)         &     (-0.59)         \\
[1em]
\shortstack{Lag\\SP500\\returns}&       0.331         &      -0.184         &      -0.176         &     -0.0587         &       0.258         &    -0.00507         &       0.157         &     -0.0580         &       0.273         &       0.450         &     -0.0878         &      -0.685         &      -0.327         &      -0.223         &      -0.258         &      -0.159         &       0.186         \\
            &      (0.73)         &     (-0.92)         &     (-0.37)         &     (-0.11)         &      (0.50)         &     (-0.01)         &      (0.27)         &     (-0.18)         &      (1.01)         &      (0.91)         &     (-0.17)         &     (-1.53)         &     (-0.70)         &     (-0.41)         &     (-0.62)         &     (-0.33)         &      (0.46)         \\
[1em]
VIX         &   0.0000797         &    0.000707\sym{*}  &   -0.000968         &    0.000413         &   -0.000873         &    0.000115         &    0.000524         &   -0.000402         &    0.000834\sym{*}  &   -0.000293         &    0.000172         &    -0.00175\sym{**} &    0.000750         &   -0.000922         &   -0.000430         &   -0.000179         &    0.000363         \\
            &      (0.13)         &      (2.40)         &     (-1.40)         &      (0.54)         &     (-1.57)         &      (0.14)         &      (0.68)         &     (-0.73)         &      (2.03)         &     (-0.64)         &      (0.28)         &     (-2.97)         &      (1.11)         &     (-1.15)         &     (-0.68)         &     (-0.30)         &      (0.47)         \\
[1em]
\shortstack{Lag\\VIX}&   -0.000614         &   -0.000143         &    0.000469         &   -0.000207         &   -0.000109         &    0.000461         &   0.0000854         &    0.000188         &   -0.000656         &   -0.000881         &    0.000762         &    0.000661         &   0.0000888         &   -0.000413         &    0.000365         &   -0.000162         &   -0.000240         \\
            &     (-1.00)         &     (-0.47)         &      (0.52)         &     (-0.27)         &     (-0.14)         &      (0.66)         &      (0.10)         &      (0.39)         &     (-1.94)         &     (-1.24)         &      (0.82)         &      (0.95)         &      (0.15)         &     (-0.56)         &      (0.59)         &     (-0.21)         &     (-0.46)         \\
[1em]
Constants   &      0.0497\sym{***}&      0.0112\sym{**} &      0.0258         &      0.0191\sym{*}  &      0.0332\sym{*}  &      0.0122         &      0.0208         &      0.0192\sym{**} &      0.0201\sym{*}  &      0.0401\sym{***}&     -0.0118         &      0.0572\sym{***}&      0.0183         &      0.0405\sym{**} &      0.0310\sym{**} &      0.0472\sym{***}&      0.0206         \\
            &      (4.49)         &      (2.77)         &      (1.58)         &      (2.01)         &      (2.35)         &      (0.83)         &      (1.82)         &      (2.85)         &      (2.46)         &      (3.95)         &     (-0.74)         &      (4.37)         &      (1.43)         &      (3.36)         &      (3.03)         &      (3.60)         &      (1.84)         \\
\hline
\(N\)       &          95         &          95         &          95         &          95         &          95         &          95         &          95         &          95         &          95         &          95         &          95         &          95         &          95         &          95         &          95         &          95         &          95         \\
\hline\hline
\multicolumn{18}{l}{\footnotesize \textit{t} statistics in parentheses}\\
\multicolumn{18}{l}{\footnotesize \sym{*} \(p<0.05\), \sym{**} \(p<0.01\), \sym{***} \(p<0.001\)}\\
\end{tabular}
}

%% file: ar_q_control_mod.tex
{
\def\sym#1{\ifmmode^{#1}\else\(^{#1}\)\fi}
\begin{tabular}{l a b a b a b a b a b a b a b a b a b a b a b a b a b a b a}
\hline\hline
            &\multicolumn{1}{c}{(1)}&\multicolumn{1}{c}{(2)}&\multicolumn{1}{c}{(3)}&\multicolumn{1}{c}{(4)}&\multicolumn{1}{c}{(5)}&\multicolumn{1}{c}{(6)}&\multicolumn{1}{c}{(7)}&\multicolumn{1}{c}{(8)}&\multicolumn{1}{c}{(9)}&\multicolumn{1}{c}{(10)}&\multicolumn{1}{c}{(11)}&\multicolumn{1}{c}{(12)}&\multicolumn{1}{c}{(13)}&\multicolumn{1}{c}{(14)}&\multicolumn{1}{c}{(15)}\\
            &\multicolumn{1}{c}{Labor}&\multicolumn{1}{c}{Forecast}&\multicolumn{1}{c}{Supervision}&\multicolumn{1}{c}{Housing}&\multicolumn{1}{c}{Payment}&\multicolumn{1}{c}{\shortstack{Small\\business}}&\multicolumn{1}{c}{Innovation}&\multicolumn{1}{c}{Regulation}&\multicolumn{1}{c}{Spending}&\multicolumn{1}{c}{\shortstack{Financial\\markets}}&\multicolumn{1}{c}{FED}&\multicolumn{1}{c}{Community}&\multicolumn{1}{c}{Compliance}&\multicolumn{1}{c}{Trade}&\multicolumn{1}{c}{Treasury}\\
\hline
1 Lag       &       0.672\sym{***}&       0.171         &       0.241         &       0.797\sym{***}&      0.0876         &       0.132         &       0.640\sym{***}&       0.489\sym{***}&       0.464\sym{***}&       0.402\sym{***}&       0.579\sym{***}&       0.277\sym{*}  &       0.567\sym{***}&       0.226\sym{*}  &       0.519\sym{***}\\
            &      (7.68)         &      (1.60)         &      (1.87)         &      (7.95)         &      (0.70)         &      (1.13)         &      (6.63)         &      (4.26)         &      (3.91)         &      (3.71)         &      (5.49)         &      (2.31)         &      (5.90)         &      (2.55)         &      (5.15)         \\
[1em]
Inflation   &       0.882         &       1.024\sym{*}  &      -0.115         &      -0.588         &      -0.224         &      0.0300         &       0.398         &      -0.266         &       0.946         &      -0.134         &      -0.707         &       0.468         &      -0.107         &      -0.141         &      -1.342\sym{**} \\
            &      (1.54)         &      (2.03)         &     (-0.15)         &     (-1.00)         &     (-0.45)         &      (0.08)         &      (0.67)         &     (-0.52)         &      (1.14)         &     (-0.23)         &     (-1.35)         &      (1.00)         &     (-0.28)         &     (-0.31)         &     (-2.90)         \\
[1em]
\shortstack{Lag\\Inflation}&      -1.284\sym{*}  &      -1.517\sym{**} &       1.844\sym{*}  &      -0.609         &      0.0924         &      -0.342         &       0.872\sym{*}  &     -0.0738         &      -0.545         &       1.087         &       0.298         &       0.753         &     -0.0731         &      -0.104         &      -0.292         \\
            &     (-2.45)         &     (-2.85)         &      (2.16)         &     (-1.61)         &      (0.25)         &     (-0.72)         &      (2.28)         &     (-0.10)         &     (-0.57)         &      (1.68)         &      (0.35)         &      (1.54)         &     (-0.17)         &     (-0.27)         &     (-0.71)         \\
[1em]
\shortstack{1 year\\bond}&       0.404         &       0.915         &       1.204         &       1.032         &       2.253         &       1.403         &       1.162         &       0.288         &       2.605         &      -1.660         &      -3.243\sym{*}  &      -2.436\sym{***}&     -0.0541         &     -0.0578         &      -1.456         \\
            &      (0.45)         &      (0.89)         &      (0.72)         &      (1.38)         &      (1.76)         &      (1.51)         &      (0.88)         &      (0.33)         &      (1.24)         &     (-1.80)         &     (-2.08)         &     (-3.69)         &     (-0.08)         &     (-0.09)         &     (-1.84)         \\
[1em]
\shortstack{Lag\\1 year\\bond}&      -2.348\sym{**} &       0.341         &       2.363         &      -0.295         &       2.467         &      -1.662         &       1.556         &      -2.346\sym{*}  &      -0.589         &       0.891         &       1.720         &      -0.474         &      -0.527         &       1.327         &      -0.119         \\
            &     (-2.87)         &      (0.38)         &      (1.32)         &     (-0.51)         &      (1.79)         &     (-1.98)         &      (1.51)         &     (-2.04)         &     (-0.21)         &      (0.91)         &      (1.33)         &     (-0.55)         &     (-0.84)         &      (1.90)         &     (-0.14)         \\
[1em]
\shortstack{SP500\\returns}&      -1.454\sym{*}  &       0.569         &       0.862         &       0.330         &      -2.589\sym{**} &      -0.235         &      -0.246         &      -0.794         &      0.0402         &       0.303         &       1.481         &       0.693         &       0.182         &     -0.0307         &       0.499         \\
            &     (-2.34)         &      (0.61)         &      (0.76)         &      (0.55)         &     (-2.88)         &     (-0.27)         &     (-0.36)         &     (-1.10)         &      (0.04)         &      (0.49)         &      (1.25)         &      (0.99)         &      (0.39)         &     (-0.06)         &      (0.77)         \\
[1em]
\shortstack{Lag\\SP500\\returns}&       0.112         &       0.135         &      -0.125         &      -0.338         &       0.284         &      -0.141         &       0.210         &       0.779         &       0.130         &      -0.410         &      -0.438         &      -0.188         &      -0.337         &      -0.312         &       0.395         \\
            &      (0.15)         &      (0.17)         &     (-0.12)         &     (-0.60)         &      (0.31)         &     (-0.18)         &      (0.27)         &      (0.92)         &      (0.08)         &     (-0.73)         &     (-0.46)         &     (-0.30)         &     (-0.61)         &     (-0.54)         &      (0.59)         \\
[1em]
VIX         &     0.00232\sym{*}  &    -0.00102         &    -0.00154         &   -0.000343         &     0.00362\sym{**} &    0.000210         &   -0.000281         &    0.000446         &   -0.000613         &    0.000131         &    -0.00170         &   -0.000469         &   -0.000563         &   -0.000418         &   -0.000140         \\
            &      (2.57)         &     (-0.82)         &     (-0.97)         &     (-0.43)         &      (3.04)         &      (0.15)         &     (-0.31)         &      (0.45)         &     (-0.40)         &      (0.17)         &     (-1.28)         &     (-0.52)         &     (-0.80)         &     (-0.62)         &     (-0.18)         \\
[1em]
\shortstack{Lag\\VIX}&   -0.000877         &    -0.00135         &    0.000293         &    0.000483         &   -0.000948         &    0.000859         &   -0.000203         &   -0.000150         &   -0.000193         &    0.000230         &     0.00127         &   -0.000238         &    0.000594         &   -0.000314         &    0.000576         \\
            &     (-0.70)         &     (-1.20)         &      (0.18)         &      (0.55)         &     (-0.74)         &      (0.75)         &     (-0.18)         &     (-0.12)         &     (-0.10)         &      (0.26)         &      (0.67)         &     (-0.27)         &      (0.60)         &     (-0.39)         &      (0.56)         \\
[1em]
Constants   &      0.0481         &       0.114\sym{***}&      0.0385         &     0.00583         &     0.00329         &      0.0230         &      0.0153         &      0.0410\sym{*}  &      0.0528\sym{*}  &      0.0373\sym{*}  &      0.0263         &      0.0689\sym{**} &      0.0277         &      0.0594\sym{***}&     0.00835         \\
            &      (1.97)         &      (5.30)         &      (1.57)         &      (0.48)         &      (0.18)         &      (1.24)         &      (1.06)         &      (2.44)         &      (2.24)         &      (2.08)         &      (0.86)         &      (3.30)         &      (1.86)         &      (4.00)         &      (0.53)         \\
\hline
\(N\)       &          95         &          95         &          95         &          95         &          95         &          95         &          95         &          95         &          95         &          95         &          95         &          95         &          95         &          95         &          95         \\
\hline\hline
\multicolumn{16}{l}{\footnotesize \textit{t} statistics in parentheses}\\
\multicolumn{16}{l}{\footnotesize \sym{*} \(p<0.05\), \sym{**} \(p<0.01\), \sym{***} \(p<0.001\)}\\
\end{tabular}
}

%% file: ms.bib
@article{brainard1967,
 ISSN = {00028282},
 URL = {http://www.jstor.org/stable/1821642},
 author = {William C. Brainard},
 journal = {The American Economic Review},
 number = {2},
 pages = {411--425},
 publisher = {American Economic Association},
 title = {Uncertainty and the Effectiveness of Policy},
 volume = {57},
 year = {1967}
}

@article{barthes1975,
 ISSN = {00286087, 1080661X},
 URL = {http://www.jstor.org/stable/468419},
 author = {Roland Barthes and Lionel Duisit},
 journal = {New Literary History},
 number = {2},
 pages = {237--272},
 publisher = {Johns Hopkins University Press},
 title = {An Introduction to the Structural Analysis of Narrative},
 volume = {6},
 year = {1975}
}

@article{bruner1991,
 ISSN = {00931896, 15397858},
 URL = {http://www.jstor.org/stable/1343711},
 author = {Jerome Bruner},
 journal = {Critical Inquiry},
 number = {1},
 pages = {1--21},
 publisher = {The University of Chicago Press},
 title = {The Narrative Construction of Reality},
 volume = {18},
 year = {1991}
}

@article{nyman2018,
author = {Nyman, Rickard and Kapadia, Sujit and Tuckett, David and Gregory, David and Ormerod, Paul and Smith, Robert},
year = {2018},
month = {01},
pages = {},
title = {News and Narratives in Financial Systems: Exploiting Big Data for Systemic Risk Assessment},
journal = {SSRN Electronic Journal},
doi = {10.2139/ssrn.3135262}
}

@book{mitchell1981,
author = {Mitchell, W. J. T.},
year = {1981},
address = {Chicago ;},
booktitle = {On narrative},
isbn = {0226532178},
language = {eng},
publisher = {University of Chicago Press},
series = {Phoenix book},
title = {On narrative },
}

@book{sarbin1986,
  title={Narrative psychology: The storied nature of human conduct.},
  author={Sarbin, Theodore R},
  year={1986},
  publisher={Praeger Publishers/Greenwood Publishing Group}
}

@article{shiller2017,
Author = {Shiller, Robert J.},
Title = {Narrative Economics},
Journal = {American Economic Review},
Volume = {107},
Number = {4},
Year = {2017},
Month = {April},
Pages = {967-1004},
DOI = {10.1257/aer.107.4.967},
URL = {http://www.aeaweb.org/articles?id=10.1257/aer.107.4.967}
}

@book{bird2009,
    title = "Natural Language Processing with Python",
    author = "Bird, Steven and Loper, Edward and Klein Ewan",
    year = "2009",
    publisher = "O'Reilly Media Inc."
}

@unpublished{mccallum2002,
      author = "Andrew Kachites McCallum",
      title = "MALLET: A Machine Learning for Language Toolkit",
      note = "{\url{http://mallet.cs.umass.edu}}",
      year = 2002}

@inproceedings{rehurek2010,
    title = {{Software Framework for Topic Modelling with Large Corpora}},
    author = {Radim {\v R}eh{\r u}{\v r}ek and Petr Sojka},
    booktitle = {{Proceedings of the LREC 2010 Workshop on New
        Challenges for NLP Frameworks}},
    pages = {45--50},
    year = 2010,
    month = May,
    day = 22,
    publisher = {ELRA},
    address = {Valletta, Malta},
    note={\url{http://is.muni.cz/publication/884893/en}},
    language={English}
}

@TechReport{eb92,
    title = {{Standard Eurobarometer (EB 92) -- Public opinions in the European Union}},
    year = {2019},
    month = {nov},
    author = {{European Commission}}
}

@TechReport{cbdc1,
    title = {{Central bank digital currencies: foundational principles and core features}},
    year = {2020},
    month = {oct},
    author = {{Bank for International Settlements}}
}

@article{amato2002,
    author = {Amato, Jeffery D. and Morris, Stephen and Shin, Hyun Song},
    title = "{Communication and Monetary Policy}",
    journal = {Oxford Review of Economic Policy},
    volume = {18},
    number = {4},
    pages = {495-503},
    year = {2002},
    month = {12},
    issn = {0266-903X},
    doi = {10.1093/oxrep/18.4.495},
    url = {https://doi.org/10.1093/oxrep/18.4.495},
    eprint = {https://academic.oup.com/oxrep/article-pdf/18/4/495/1210884/180495.pdf},
}

@article{andersson2006,
title = "Monetary policy signaling and movements in the term structure of interest rates",
journal = "Journal of Monetary Economics",
volume = "53",
number = "8",
pages = "1815 - 1855",
year = "2006",
issn = "0304-3932",
doi = "https://doi.org/10.1016/j.jmoneco.2006.06.002",
url = "http://www.sciencedirect.com/science/article/pii/S0304393206001346",
author = "Malin Andersson and Hans Dillén and Peter Sellin"
}

@article{armelius2020,
title = {Spread the Word: International spillovers from central bank communication},
journal = {Journal of International Money and Finance},
volume = {103},
pages = {102116},
year = {2020},
issn = {0261-5606},
doi = {https://doi.org/10.1016/j.jimonfin.2019.102116},
url = {https://www.sciencedirect.com/science/article/pii/S0261560619302967},
author = {Hanna Armelius and Christoph Bertsch and Isaiah Hull and Xin Zhang}
}

@article{blinder2008,
 ISSN = {00220515},
 URL = {http://www.jstor.org/stable/27647085},
 author = {Alan S. Blinder and Michael Ehrmann and Marcel Fratzscher and Jakob de Haan and David-Jan Jansen},
 journal = {Journal of Economic Literature},
 number = {4},
 pages = {910--945},
 publisher = {American Economic Association},
 title = {Central Bank Communication and Monetary Policy: A Survey of Theory and Evidence},
 volume = {46},
 year = {2008}
}

@article{blinder2018,
 ISSN = {25740768, 25740776},
 URL = {https://www.jstor.org/stable/26452803},
 author = {Alan S. Blinder},
 journal = {AEA Papers and Proceedings},
 pages = {567--571},
 publisher = {American Economic Association},
 title = {Through a Crystal Ball Darkly: The Future of Monetary Policy Communication},
 volume = {108},
 year = {2018}
}

@article{born2014,
 ISSN = {00130133, 14680297},
 URL = {http://www.jstor.org/stable/42919216},
 author = {Benjamin Born and Michael Ehrmann and Marcel Fratzscher},
 journal = {The Economic Journal},
 number = {577},
 pages = {701--734},
 publisher = {Wiley},
 title = {Central Bank Communication on Financial Stability},
 volume = {124},
 year = {2014}
}

@article{cochrane2002,
 ISSN = {00028282},
 URL = {http://www.jstor.org/stable/3083383},
 author = {John H. Cochrane and Monika Piazzesi},
 journal = {The American Economic Review},
 number = {2},
 pages = {90--95},
 publisher = {American Economic Association},
 title = {The Fed and Interest Rates: A High-Frequency Identification},
 volume = {92},
 year = {2002}
}

@article{coibion2020,
title = "Inflation expectations as a policy tool?",
journal = "Journal of International Economics",
volume = "124",
pages = "103297",
year = "2020",
note = "NBER International Seminar on Macroeconomics 2019",
issn = "0022-1996",
doi = "https://doi.org/10.1016/j.jinteco.2020.103297",
url = "http://www.sciencedirect.com/science/article/pii/S0022199620300167",
author = "Olivier Coibion and Yuriy Gorodnichenko and Saten Kumar and Mathieu Pedemonte"
}

@article{cook1989,
title = "The effect of changes in the federal funds rate target on market interest rates in the 1970s",
journal = "Journal of Monetary Economics",
volume = "24",
number = "3",
pages = "331 - 351",
year = "1989",
issn = "0304-3932",
doi = "https://doi.org/10.1016/0304-3932(89)90025-1",
url = "http://www.sciencedirect.com/science/article/pii/0304393289900251",
author = "Timothy Cook and Thomas Hahn"
}

@article{ellen2019,
  author={Saskia ter Ellen and Vegard H. Larsen and Leif Anders Thorsrud},
  title="Narrative monetary policy surprises and the media",
  year="2019",
  type={Working Papers},
  url={https://ideas.repec.org/p/bny/wpaper/0078.html},
  number={No 06/2019},
}

@article{haldane2018,
Author = {Haldane, Andrew and McMahon, Michael},
Title = {Central Bank Communications and the General Public},
Journal = {AEA Papers and Proceedings},
Volume = {108},
Year = {2018},
Month = {May},
Pages = {578-83},
DOI = {10.1257/pandp.20181082},
URL = {http://www.aeaweb.org/articles?id=10.1257/pandp.20181082}}

@article{hansen2019,
title = "The long-run information effect of central bank communication",
journal = "Journal of Monetary Economics",
volume = "108",
pages = "185 - 202",
year = "2019",
note = "“Central Bank Communications:From Mystery to Transparency”May 23-24, 2019Annual Research Conference ofthe National Bank of UkraineOrganized in cooperation withNarodowy Bank Polski",
issn = "0304-3932",
doi = "https://doi.org/10.1016/j.jmoneco.2019.09.002",
url = "http://www.sciencedirect.com/science/article/pii/S0304393219301606",
author = "Stephen Hansen and Michael McMahon and Matthew Tong"
}

@article{hansen2017,
    author = {Hansen, Stephen and McMahon, Michael and Prat, Andrea},
    title = "{Transparency and Deliberation Within the FOMC: A Computational Linguistics Approach*}",
    journal = {The Quarterly Journal of Economics},
    volume = {133},
    number = {2},
    pages = {801-870},
    year = {2017},
    month = {10},
    issn = {0033-5533},
    doi = {10.1093/qje/qjx045},
    url = {https://doi.org/10.1093/qje/qjx045},
    eprint = {https://academic.oup.com/qje/article-pdf/133/2/801/24518498/qjx045.pdf},
}

@article{hayo2014,
title = "The German public and its trust in the ECB: The role of knowledge and information search",
journal = "Journal of International Money and Finance",
volume = "47",
pages = "286 - 303",
year = "2014",
issn = "0261-5606",
doi = "https://doi.org/10.1016/j.jimonfin.2014.07.003",
url = "http://www.sciencedirect.com/science/article/pii/S026156061400117X",
author = "Bernd Hayo and Edith Neuenkirch"
}

@article{jansen2005,
title = "Talking heads: the effects of ECB statements on the euro–dollar exchange rate",
journal = "Journal of International Money and Finance",
volume = "24",
number = "2",
pages = "343 - 361",
year = "2005",
note = "Exchange Rate Economics",
issn = "0261-5606",
doi = "https://doi.org/10.1016/j.jimonfin.2004.12.009",
url = "http://www.sciencedirect.com/science/article/pii/S0261560604001202",
author = "David-Jan Jansen and Jakob De Haan"
}

@article{kumar2015,
 ISSN = {00072303, 15334465},
 URL = {http://www.jstor.org/stable/43752171},
 author = {Saten Kumar and Olivier Coibion and Hassan Afrouzi and Yuriy Gorodnichenko},
 journal = {Brookings Papers on Economic Activity},
 pages = {151--208},
 publisher = {Brookings Institution Press},
 title = {Inflation Targeting Does Not Anchor Inflation Expectations: Evidence from Firms in New Zealand},
 year = {2015}
}

@article{lamar2019,
title = "Central bank announcements: Big news for little people?",
journal = "Journal of Monetary Economics",
volume = "108",
pages = "21 - 38",
year = "2019",
note = "“Central Bank Communications:From Mystery to Transparency”May 23-24, 2019Annual Research Conference ofthe National Bank of UkraineOrganized in cooperation withNarodowy Bank Polski",
issn = "0304-3932",
doi = "https://doi.org/10.1016/j.jmoneco.2019.08.014",
url = "http://www.sciencedirect.com/science/article/pii/S0304393219301473",
author = "Michael J. Lamla and Dmitri V. Vinogradov"
}

@article{nakamura2018,
    author = {Nakamura, Emi and Steinsson, Jón},
    title = "{High-Frequency Identification of Monetary Non-Neutrality: The Information Effect*}",
    journal = {The Quarterly Journal of Economics},
    volume = {133},
    number = {3},
    pages = {1283-1330},
    year = {2018},
    month = {01},

    issn = {0033-5533},
    doi = {10.1093/qje/qjy004},
    url = {https://doi.org/10.1093/qje/qjy004},
    eprint = {https://academic.oup.com/qje/article-pdf/133/3/1283/25112639/qjy004.pdf},
}

@article{woodford2001,
  author={Michael Woodford},
  title={{Monetary policy in the information economy}},
  journal={Proceedings - Economic Policy Symposium - Jackson Hole},
  year=2001,
  volume={},
  number={},
  pages={297-370},
  month={},
  doi={},
  url={https://ideas.repec.org/a/fip/fedkpr/y2001p297-370.html}
}

@techreport{woodford2005,
 title = "Central Bank Communication and Policy Effectiveness",
 author = "Woodford, Michael",
 institution = "National Bureau of Economic Research",
 type = "Working Paper",
 series = "Working Paper Series",
 number = "11898",
 year = "2005",
 month = "December",
 doi = {10.3386/w11898},
 URL = "http://www.nber.org/papers/w11898",
}

@article{newey1987,
issn = {0012-9682},
journal = {Econometrica},
pages = {703},
volume = {55},
publisher = {Blackwell Publishers Ltd.},
year = {1987},
title = {A simple, positive semi-definite, heteroskedasticity and autocorrelation consistent covariance matrix},
author = {Newey, Whitney K. and West, Kenneth D.},
}

@article{blei2003,
    author = {Blei, David and Ng, Andrew and Jordan, Michael},
    year = {2003},
    month = {03},
    pages = {993-1022},
    title = {Latent Dirichlet Allocation},
    volume = {3},
    journal = {The Journal of Machine Learning Research}
}

@article{blei2006,
author = {Blei, David and Lafferty, John},
title = {Dynamic Topic Models},
year = {2006},
isbn = {1595933832},
publisher = {Association for Computing Machinery},
address = {New York, NY, USA},
url = {https://doi.org/10.1145/1143844.1143859},
doi = {10.1145/1143844.1143859},
booktitle = {Proceedings of the 23rd International Conference on Machine Learning},
pages = {113–120},
numpages = {8},
location = {Pittsburgh, Pennsylvania, USA},
series = {ICML ’06}
}

@article{blei2007,
    author = {Blei, David and Lafferty, John},
    year = {2007},
    month = {08},
    pages = {},
    title = {A correlated topic model of Science},
    volume = {1},
    journal = {The Annals of Applied Statistics},
    doi = {10.1214/07-AOAS114}
}

@article{deerwester1990,
author={Deerwester,Scott and Dumais,Susan T. and Furnas,George W. and Landauer,Thomas K. and Harshman,Richard},
year={1990},
month={09},
title={Indexing by Latent Semantic Analysis},
journal={Journal of the American Society for Information Science (1986-1998)},
volume={41},
number={6},
pages={391},
note={Copyright - Copyright Wiley Periodicals Inc. Sep 1990; Last updated - 2019-11-23; SubjectsTermNotLitGenreText - New York},
isbn={00028231},
language={en},
url={https://search-proquest-com.ezproxy.ub.gu.se/docview/216891549?accountid=11162},
}

@article{gentzkow2019,
issn = {0022-0515},
journal = {Journal of Economic Literature},
pages = {535--574},
volume = {57},
publisher = {American Economic Association},
number = {3},
year = {2019},
title = {Text as Data},
language = {eng},
author = {Gentzkow, Matthew and Kelly, Bryan and Taddy, Matt},
}

@article {griffiths2004,
	author = {Griffiths, Thomas L. and Steyvers, Mark},
	title = {Finding scientific topics},
	volume = {101},
	number = {suppl 1},
	pages = {5228--5235},
	year = {2004},
	doi = {10.1073/pnas.0307752101},
	publisher = {National Academy of Sciences},
	issn = {0027-8424},
	URL = {https://www.pnas.org/content/101/suppl_1/5228},
	eprint = {https://www.pnas.org/content/101/suppl_1/5228.full.pdf},
	journal = {Proceedings of the National Academy of Sciences}
}

@inproceedings{hofmann1999,
    title={Probabilistic latent semantic indexing},
    author={Thomas Hofmann},
    booktitle={SIGIR '99},
    year={1999}
}

@inproceedings{newman2010,
author = {Newman, David and Lau, Jey and Grieser, Karl and Baldwin, Timothy},
year = {2010},
month = {01},
pages = {100-108},
title = {Automatic Evaluation of Topic Coherence.},
journal = {Human Language Technologies: The 2010 Annual Conference of the North American Chapter of the ACL}
}

@MISC{papadimitriou1997,
    author = {Christos H. Papadimitriou and Prabhakar Raghavan and Hisao Tamaki and Santosh
            Vempala},
    title = {Latent Semantic Indexing: A Probabilistic Analysis},
    year = {1997}
}

@inproceedings{roder2015,
author = {R\"{o}der, Michael and Both, Andreas and Hinneburg, Alexander},
title = {Exploring the Space of Topic Coherence Measures},
year = {2015},
isbn = {9781450333177},
publisher = {Association for Computing Machinery},
address = {New York, NY, USA},
url = {https://doi.org/10.1145/2684822.2685324},
doi = {10.1145/2684822.2685324},
booktitle = {Proceedings of the Eighth ACM International Conference on Web Search and Data Mining},
pages = {399–408},
numpages = {10},
location = {Shanghai, China},
series = {WSDM ’15}
}

@inproceedings{teh2004,
author = {Teh, Yee Whye and Jordan, Michael I. and Beal, Matthew J. and Blei, David M.},
title = {Sharing Clusters among Related Groups: Hierarchical Dirichlet Processes},
year = {2004},
publisher = {MIT Press},
address = {Cambridge, MA, USA},
booktitle = {Proceedings of the 17th International Conference on Neural Information Processing Systems},
pages = {1385–1392},
numpages = {8},
location = {Vancouver, British Columbia, Canada},
series = {NIPS’04}
}

@inproceedings{wallach2009,
author = {Wallach, Hanna M. and Murray, Iain and Salakhutdinov, Ruslan and Mimno, David},
title = {Evaluation Methods for Topic Models},
year = {2009},
isbn = {9781605585161},
publisher = {Association for Computing Machinery},
address = {New York, NY, USA},
url = {https://doi.org/10.1145/1553374.1553515},
doi = {10.1145/1553374.1553515},
booktitle = {Proceedings of the 26th Annual International Conference on Machine Learning},
pages = {1105–1112},
numpages = {8},
location = {Montreal, Quebec, Canada},
series = {ICML ’09}
}
